\documentstyle[prb,aps,epsf,eqsecnum]{revtex}
\newcommand{\lapprox}{\stackrel{<}{\sim}}
\newcommand{\gapprox}{\stackrel{>}{\sim}}
\begin{document}
\draft
\twocolumn[\hsize\textwidth\columnwidth\hsize\csname  
@twocolumnfalse\endcsname
\author{P. Pieri and G.C. Strinati}
\address{Dipartimento di Matematica e Fisica, Sezione INFM, 
Universit\`{a} di Camerino, I-62032 Camerino, Italy\\
e-mail: pieri@str.unicam.it; strinati@camars.unicam.it}
\date{October 14, 1998}

\title{On the correct strong-coupling limit in the evolution from BCS
superconductivity to Bose-Einstein condensation}

\maketitle
\hspace*{-0.25ex}\begin{abstract}
We consider the problem of the crossover from BCS superconductivity 
to Bose-Einstein condensation in three dimensions for a system of fermions with an attractive 
interaction, for which we adopt the simplifying assumption of a suitably regularized 
point-contact interaction.
We examine in a critical way the fermionic (self-consistent) T-matrix approximation
which has been widely utilized in the literature to describe this crossover
\emph{above\/} the superconducting critical temperature, 
and show that it fails to yield the correct behaviour of the system in the 
strong-coupling limit, where composite bosons form as tightly bound fermion pairs.
We then set up the correct approximation for a ``dilute'' system of 
composite bosons
and show that an entire new class of diagrams has to be considered in the 
place of 
the fermionic T-matrix approximation for the self-energy.
This new class of diagrams correctly describes \emph{both\/} the weak- and 
strong-coupling limits, and consequently results into an improved 
interpolation scheme for the intermediate (crossover) region.
In this context, we provide also a systematic mapping between the corresponding
diagrammatic theories for the composite bosons and the constituent fermions.
As a preliminary result to demonstrate the numerical effect of our new class 
of diagrams on physical quantities, we calculate the value of the scattering 
length
for composite bosons in the strong-coupling limit and show that it is considerably
modified with respect to the result obtained within the self-consistent fermionic 
T-matrix approximation.
\end{abstract}
\hspace*{-0.25ex}
]

\narrowtext

\section{Introduction}

The problem of the crossover from BCS superconductivity to Bose-Einstein (BE) 
condensation has attracted considerable interest lately,\cite{Uemura} and especially 
after the recent ARPES experiments in cuprate superconductors which have shown the 
existence of a (pseudo) gap at temperatures above the superconducting transition 
temperature $T_{c}$.\cite{Ding,Loeser}
This observation has prompted the proposal by many authors of the possible presence
of quasi-bound fermionic pairs above $T_{c}$ and up to a second temperature scale
$T^{*}$.
More generally, one may think of ascribing the anomalous experimental properties of
the cuprate superconductors in the normal state (i.e., for temperatures $T$ such that
$T_{c} \lapprox T \lapprox T^{*}$) as well as the possible break-down of the
Fermi-liquid theory to the presence of these quasi-bound pairs.\cite{Randeria-97-2}

From the theoretical point of view, this crossover problem (as well as related
crossover problems, like the one associated with the Mott transition 
\cite{Janis}) poses a compelling challenge, because approximations that are 
valid on the one side of the crossover are not necessarily valid on the opposite side.
In addition, the crossover region is characterized (or even defined) by the absence 
of a ``small'' parameter, which would allow one to control the approximations.
For systems that are sufficiently ``dilute'' (such that, for instance, for given 
strength of the inter-particle interaction, the density $\rho$ can be taken to be 
arbitrarily small), one could directly exploit the well-known results obtained for 
``dilute'' fermionic systems with $\rho_{F} = \rho$, on the one hand,\cite{Galitskii}
or for ``dilute'' bosonic systems with $\rho_{B} = \rho/2$, on the other 
hand,\cite{Beliaev} to get a correct description of the limits on the two sides 
of the crossover.

For a ``dilute'' fermionic system the two-body equation plays an essential
role.\cite{Galitskii}
In particular, in three dimensions the low-energy two-body scattering process can 
be parametrized in terms of the \emph{scattering length\/} $a_{F}$, which is negative 
for weak coupling and positive for strong coupling (i.e., in the presence of a bound 
state for the attractive interaction), and diverges when the coupling strength
suffices for the bound state to appear.
For strong coupling, $a_{F}$ gives the size of the bound state.
The many-body approach to a ``dilute'' fermionic system then introduces the small 
dimensionless parameter $k_{F}a_{F}$ to select the relevant diagrammatic contributions, 
whereby the Fermi wave vector identifies the average interparticle distance 
$k_{F}^{-1}$ and $a_{F}$ emerges directly from the many-particle T-matrix 
(which is defined by repeated scattering in the particle-particle channel at 
finite fermionic density).
In the original treatment with a purely \emph{repulsive\/} interaction, no bound state 
occurred and $a_{F}$ remained finite, in such a way that the parameter $k_{F}a_{F}$ 
could unambiguously be taken to be much less than unity.\cite{Galitskii}
In the case of interest to us of an \emph{attractive\/} interaction, which develops
a bound state in three dimensions, however, the parameter $k_{F}|a_{F}|$ may readily 
exceed unity when the strength of the attractive interaction increases toward its 
critical value for the appearance of a bound state.
In this case, the many-body T-matrix can no longer be identified with $a_{F}$.
Furthermore, past the critical interaction strength where the two-body problem 
develops a bound state (with $a_{F}$ turning again finite and eventually vanishing 
in the strong-coupling limit), the many-body T-matrix acquires a singularity (pole) 
for all interaction strengths and no identification of the T-matrix with $a_{F}$ 
is clearly possible any longer.
Therefore, although the parameter $k_{F}a_{F}$ may still serve to readily locate which 
side of the crossover one is examining (or even qualitatively, how close to the 
crossover region one is) by relying on the results of the two-body problem, the 
smallness of this parameter might not by itself be sufficient to guarantee that an 
approximation, selected for the weak-coupling limit, is still valid in the 
strong-coupling limit.

In particular, the T-matrix approximation for the fermionic self-energy (which has 
invariably been regarded in the literature as representing \emph{the\/} 
dilute-approach approximation to the BCS-BE crossover 
problem)\cite{Fresard,Haussmann,Micnas,Levin-97,Randeria-97-1,KKK} 
is \emph{not\/} expected to remain valid as soon as the two-body bound state develops, 
since for this approximation the power counting in the small parameter $k_{F} a_{F}$ 
relies just on the identification of the many-body T-matrix with $a_{F}$.
This statement is consistent with the physical picture that, 
\emph{as soon as the two-body bound state develops, it is the residual interaction 
between the composite bosons to determine the ``diluteness'' condition of the system\/} 
and not the original attraction between the constituent fermions, which produces the 
bound state to begin with and relatively to which the system is in the strong-coupling 
limit.\cite{footnote-1}
It is then clear that a correct description of the strong-coupling limit can be
obtained only by selecting the relevant approximations directly for a \emph{dilute 
system of composite bosons\/}, rather than relying on the fermionic T-matrix 
approximation, which is valid by construction for a \emph{dilute system of fermions\/}.

From previous work on the functional-integral approach to the crossover from BCS to 
BE,\cite{Randeria-93,PS-96,Zwerger} one knows that approximations (like the BCS mean 
field at zero temperature), which give a satisfactory account of the weak-coupling 
limit, become inadequate in the strong-coupling limit, and that only by including 
fluctuation corrections at least at the one-loop level a sensible description of 
the effective bosonic system in the strong-coupling limit results.
This remark has actually suggested to deal with the crossover problem 
\emph{in reverse\/},\cite{PS-96} that is, by first envisaging approximations which give 
a satisfactory description of the strong-coupling (bosonic) limit and by extrapolating
them toward the weak-coupling (fermionic) limit, where they are expected to
work properly as well.\cite{footnote-Marini}

One then anticipates the set of many-body (self-energy) diagrams, which describe 
the bosonic limit, to be much richer than the corresponding set of diagrams, which 
describe the fermionic limit.
In fact, from the previous discussion we do \emph{not\/} expect the fermionic 
T-matrix approximation to be appropriate for describing the ``dilute'' bosonic 
limit properly.
We will indeed show below that the fermionic T-matrix approximation corresponds in 
the strong-coupling limit to the standard bosonic Hartree-Fock approximation, and
that, for this reason, it misses all but one of the infinite set of self-energy 
diagrams associated with a ``dilute'' bosonic system.\cite{footnote-2}

The dynamics of a ``dilute'' system of \emph{true\/} (point-like) bosons can be 
accounted for by the corresponding T-matrix approximation (aside from a narrow region 
about $T_{c}$).\cite{Beliaev,Popov-1,Popov-2}
Using a representation with a symmetrized two-body interaction, the diagrams 
corresponding to the bosonic T-matrix approximation contain a sequence of pairs of 
propagators running in the same direction, with only one additional propagator running 
in the opposite direction (see Fig.~\ref{fig:bosloop}b of subsection III-A below). 
The first term of this sequence (i.e., the one with no pairs but only with the 
single propagator running backward) corresponds to the standard bosonic 
Hartree-Fock approximation.\cite{footnote-2} 
Since \emph{all\/} other terms (with an arbitrary number of pairs running in the same 
direction) are of the \emph{same order\/} in the ``gas parameter'' 
$\rho_{B}^{1/3} a_{B}$ (where $a_{B}$ is the (positive) scattering length for the 
low-energy two-boson scattering problem), keeping the Hartree-Fock term only is 
evidently not justified for a ``dilute'' bosonic system.

On physical grounds, for a ``dilute'' system of \emph{composite\/} bosons (obtained
by pairing fermions via the attractive interaction in the strong-coupling limit)
one expects the same picture to emerge (as it will also be confirmed by the results 
of the present paper).
For this system, the bosonic single-particle propagator corresponds to a 
\emph{two-fermion Green's function\/} in the particle-particle channel (with opposite 
spins for a spinless boson), the analogue of the bosonic self-energy corrections 
being associated in the fermionic particle-particle channel with insertions of 
the many-body effective interaction.
The complete \emph{correspondence\/} between the two (bosonic and fermionic) 
diagrammatic structures requires one to establish a well-defined mapping between 
the building blocks of the two structures (namely, propagators and interaction
vertices).
One of the purposes of this paper is to discuss this mapping between the two 
diagrammatic structures, thus complementing the mapping established by means of 
functional integrals in Ref.\ 16.
It is worth mentioning in this respect that the correspondence we shall find between 
the \emph{symmetry factors\/} of the bosonic diagrammatic structure, on the one hand,
and the number of independent diagrams in the associated fermionic structure, 
on the other hand, provides a compelling check on our mapping procedure. 

The reason for choosing the particle-particle fermionic channel in order to single 
out the (approximate) dynamics of the system in the strong-coupling limit stems 
from the obvious fact that this channel only has \emph{physical relevance\/} in this 
limit.
The single-particle fermionic propagator (or, equivalently, the fermionic 
irreducible self-energy) is then obtained as a \emph{derived\/} quantity, by suitably
closing the two-fermion diagrams with a single fermionic line.
In this way, the Baym-Kadanoff procedure \cite{BK,Baym} is effectively inverted, 
since we construct the one-particle (fermionic) self-energy diagrams in order to 
recover the \emph{chosen set\/} of two-particle diagrams via functional
differentiation, rather than following the standard procedure of deriving 
the two-particle diagrams from a chosen set of one-particle diagrams.\cite{footnote-3}

As already mentioned,\cite{footnote-1} the interaction between the composite
bosons effectively vanishes in the extreme strong-coupling limit.
In this \emph{extreme\/} limit, neither the Hartree-Fock bosonic correction
(which results from the standard fermionic T-matrix) nor the set of ``low-density''
bosonic self-energy diagrams (which are included in the present approach) need then 
to be taken into account, as they both correspond to interaction corrections.
The many-body corrections introduced in this paper in the bosonic limit are thus
expected to be important whenever the interaction between the composite bosons
has direct influence on physical properties, such as the critical temperature
$T_{c}$ and the bosonic binding energy (pseudogap).\cite{KKK}
In addition, including the interaction between the composite bosons should make
our ``low-density'' approximation to be valid somewhat closer to the crossover region,
over and above the boundary set by retaining the Hartree-Fock term only.
In this respect, our finding that the composite-boson scattering lenght $a_{B}$
is \emph{smaller\/} than the value $a_{B} = 2 a_{F}$ (obtained in Ref.\ 9
within the fermionic T-matrix approximation) effectively restricts the range of 
the crossover region on the BE side, and makes the eventual extrapolation of our 
approximation through the crossover region more reliable.

In this paper, we present the formal theory for the choice of the fermionic 
self-energy diagrams starting from the strong-coupling side, for temperatures 
\emph{above\/} $T_{c}$ (in the sense that no anomalous single-particle fermionic 
propagator will be considered) and for three spatial dimensions.
The corresponding theory below $T_{c}$ remains to be developed.
An extensive numerical study based on this approximation is under way and will be
discussed separately. 
The only numerical calculation presented in this paper concerns the value of the
composite-boson scattering length $a_{B}$ in the strong-coupling limit.

The plan of the paper is as follows.
In Section II we discuss some introductory material, which is necessary
for setting up our ``low-density'' approximation for composite bosons. Specifically:
We introduce a suitable regularization for the fermionic point-contact interaction,
which allows us to select readily the relevant classes of fermionic diagrams;
We summarize the mapping onto a bosonic system in the strong-coupling limit,
obtained by the procedure of Ref.\ 16;  
We discuss the standard fermionic T-matrix approximation and show how the Hartree-Fock
approximation for composite bosons results from it in the strong-coupling limit.
In Section III, after recalling the standard theory of the ``low-density'' Bose gas for 
true (point-like) bosons, we introduce the theory of the ``low-density'' Bose gas for 
composite bosons and obtain the fermionic self-energy from the bosonic self-energy
in the ``low-density'' approximation.
In Section IV we present the numerical calculation for the composite-boson scattering 
length $a_{B}$ in the strong-coupling limit and discuss the physical implications 
of our result. 
Section V gives our conclusions.
The Appendices discuss more technical material. 
Specifically, in Appendix A we prove that, for our choice of a ``contact'' fermionic 
interaction, the composite-boson propagator coincides exactly with the fermionic 
generalized particle-particle ladder. 
In Appendix B we discuss in detail the mapping between the bosonic and fermionic 
diagrammatic structures and demonstrate how the symmetry factors of the bosonic theory 
translates onto the number of independent diagrams in the fermionic theory, by 
working out two examples in detail. 
\section{Building blocks of the diagrammatic structure for composite bosons}

In this Section we discuss the diagrammatic structure that generically describes the 
composite bosons in terms of the constituent fermions, as a preliminary step for 
setting up the ``low-density'' approximation for composite bosons in the next Section. 
Our construction rests on a judicious choice of the fermionic interaction, which 
(albeit without loss of generality) greatly reduces the number and considerably
simplifies the expressions of the Feynman diagrams to be taken into account. 
We will show how this construction fits with the general mapping established 
in Ref.\ 16 via functional-integral methods. We shall, in turn, utilize the
results of Ref.\ 16 to single out the lowest-order (four-point) bosonic 
vertex as the relevant one in the strong-coupling limit.
We shall further show that the standard fermionic T-matrix approximation 
for a ``dilute'' Fermi gas \cite{Galitskii,Haussmann} reproduces in the strong-coupling 
limit the Hartree-Fock approximation for composite bosons,\cite{footnote-2} which can 
be defined in analogy with the standard result for true (point-like) 
bosons.\cite{Popov-1,Popov-2} 
In this way, we shall make manifest the necessity of superseding the standard fermionic
T-matrix approximation when dealing with the crossover problem and of replacing it by 
a more complete approximation, which enables us instead to recover the ``low-density'' 
approximation for composite bosons in the strong-coupling limit. 

\subsection{Regularization of the fermionic interaction}

We begin by considering the following simple model Hamiltonian for interacting
fermions (we set Planck $\hbar$ and Boltzmann $k_{B}$ constants equal to unity
throughout):
 
\begin{eqnarray}
& & H  =  \sum_{\sigma} \int d{\mathbf r} \, \psi_{\sigma}^{\dagger}({\mathbf r}) \left(
- \frac{\nabla^2}{2m} - \mu \right) \psi_{\sigma}({\mathbf r})  \nonumber  \\
& & +  \frac{1}{2} \sum_{\sigma, \sigma'} \int d{\mathbf r} \, d{\mathbf r'}
  \psi_{\sigma}^{\dagger}({\mathbf r}) \psi_{\sigma'}^{\dagger}({\mathbf r'})
  V_{{\mathrm eff}}({\mathbf r}-{\mathbf r'}) 
  \psi_{\sigma'}({\mathbf r'}) \psi_{\sigma}({\mathbf r}) 
                                                         \label{Eq:Hamiltonian}
\end{eqnarray}

\noindent
where $\psi_{\sigma}({\mathbf r}$) is the fermionic field operator with spin projection 
$\sigma = (\uparrow, \downarrow$), $m$ the fermionic (effective) mass, $\mu$ the 
fermionic chemical potential, and $V_{{\mathrm eff}}({\mathbf r}-{\mathbf r'})$ an 
{\it effective potential\/} that provides the \emph{attraction\/} between fermions.
 
To simplify the ensuing many-body diagrammatic structure considerably
(and yet preserving the physical effects we are after), we adopt for $V_{{\mathrm eff}}$ 
the simple form of a ``contact'' potential \cite{footnote-contact}

\begin{equation}
V_{{\mathrm eff}}({\mathbf r}-{\mathbf r'}) = v_{o} \,\, \delta ({\mathbf r}-{\mathbf r'})
                                                              \label{Eq:deltafunc}
\end{equation}

\noindent
where $v_{o}$ is a negative constant. 
With this choice, the interaction affects only fermions with opposite spins in
the Hamiltonian (\ref{Eq:Hamiltonian}) owing to Pauli principle.
A suitable \emph{regularization\/} of the potential (\ref{Eq:deltafunc}) is, however, 
required to get accurate control of the many-body diagrammatic structure. 
In particular, the equation (in the center-of-mass frame)

\begin{equation}
\frac{m}{4 \pi a_{F}} \, = \, \frac{1}{v_{o}} \, + \, \int \! \frac{d{\mathbf k}}
{(2 \pi)^{3}} \frac{m}{{\mathbf k}^{2}}                    \label{ferm-scatt-ampl}
\end{equation}

\noindent
for the \emph{fermionic scattering length\/} $a_{F}$ associated with the 
potential (\ref{Eq:deltafunc}) is ill-defined, since the integral over the 
three-dimensional wave vector ${\mathbf k}$ is ultraviolet divergent.
The delta-function potential (\ref{Eq:deltafunc}) is then effectively  regularized, by 
introducing an ultraviolet cutoff $k_{o}$ in the integral of Eq.~(\ref{ferm-scatt-ampl}) and letting
$v_{o} \, \rightarrow \, 0$ as $k_{o} \, \rightarrow \, \infty$, in order to
keep $a_{F}$ fixed at a chosen \emph{finite\/} value. 
The required relation between $v_{o}$ and $k_{o}$ is obtained directly from 
Eq.~(\ref{ferm-scatt-ampl}). One finds:

\begin{equation}
v_{o} \, = \, - \, \frac{2 \pi^{2}}{m k_{o}} \, - \, \frac{\pi^{3}}{m a_{F} k_{o}^{2}}
                                                               \label{vo}
\end{equation}

\noindent
when $k_{o} |a_{F}| \gg 1$. Note that $a_{F}$ enters the expression (\ref{vo}) only
through the subleading term proportional to $k_{o}^{-2}$.
Keeping this subleading term turns out to be essential for our purposes.
[Note that, if $v_{o}$ were positive instead of negative, the scattering length
$a_{F}$ would vanish in the limit $k_{o} \, \rightarrow \, \infty$ even keeping
$v_{o}$ finite.]

\begin{figure}
\narrowtext 
\epsfxsize=3.3in
\epsfbox{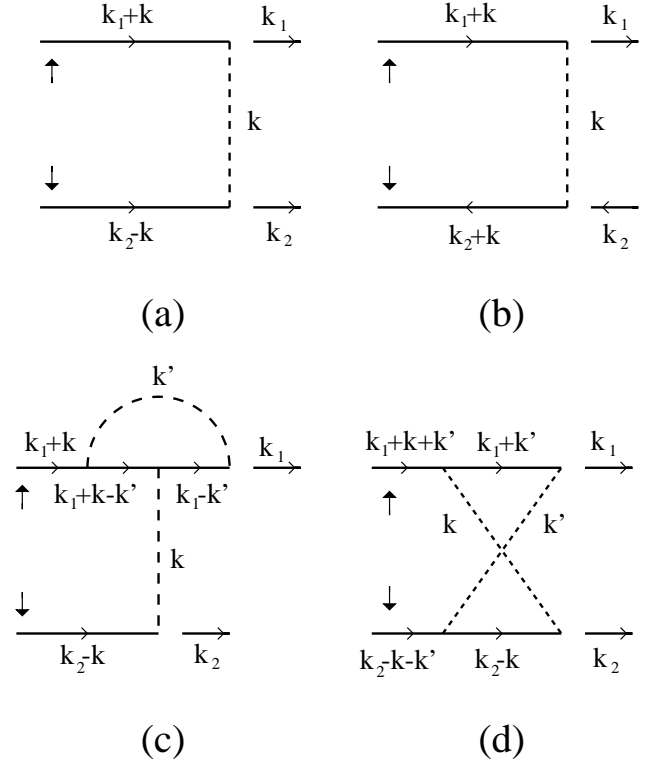}
\vspace{.2truecm}
\caption{\emph{(a)} 
Particle-particle and \emph{(b)} particle-hole rung; \emph{(c)} vertex 
correction and \emph{(d)} two-particle effective interaction. Full and broken 
lines represent the ``bare'' single-particle fermionic Green's function and 
the fermionic interaction, respectively. Four-momenta are indicated and spin 
labels are represented by up and down arrows.}
\label{fig:rung}
\end{figure}

With the regularization (\ref{vo}) for the potential, the classification of the
many-body diagrams gets considerably simplified, since only specific sub-structures 
of these diagrams survive when the limit $k_{o} \, \rightarrow \, \infty$ is
eventually taken. 
Let us consider, in particular, the characteristic rungs occurring in the theory 
of a ``dilute'' Fermi gas,\cite{Galitskii} which are depicted in Fig.~\ref{fig:rung}a
and Fig.~\ref{fig:rung}b, respectively.
For the particle-particle rung of Fig.~\ref{fig:rung}a we obtain:

\begin{eqnarray}
& &I_{pp}({\mathbf k_{1}}, \omega_{n_{1}}; {\mathbf k_{2}}, \omega_{n_{2}}) = 
\frac{1}{\beta} \sum_{\omega_{n}}\int \frac{d{\mathbf k}}{(2 \pi)^{3}}
\, v({\mathbf k})\nonumber \\
& &\times \, {\mathcal G}^{o}({\mathbf k_{1} + k}, \omega_{n_{1}} + 
\omega_{n}) \,
  {\mathcal G}^{o}({\mathbf k_{2} - k}, \omega_{n_{2}} - \omega_{n}) \nonumber \\
& &= v_{o} \int^{k_{o}} \hspace{-.3truecm}\frac{d{\mathbf k}}{(2 \pi)^{3}} \,
\frac{\left[f_{F}(\xi({\mathbf k_{1} + k})) \, + \, f_{F}(\xi({\mathbf k_{2} - k})) 
      \, - \, 1 \right]}
{i(\omega_{n_{2}} \, + \, \omega_{n_{1}}) \, - \, \xi({\mathbf k_{1} + k}) \, - \,
\xi({\mathbf k_{2} - k})}\nonumber \\
& &= v_{o} \, \left( \frac{m k_{o}}{2 \pi^{2}} \, + \, 
R_{pp}({\mathbf k_{1}}, \omega_{n_{1}}; {\mathbf k_{2}}, \omega_{n_{2}}) \right) 
\label{I-pp}
\end{eqnarray}

\noindent
where $\beta$ is the inverse temperature, $\omega_{n} = (2n + 1) \pi \beta^{-1}$
($n$ integer) a fermionic Matsubara frequency, ${\mathcal G}^{o}$ a (bare) 
single-particle fermionic Green's function, 
$f_{F}(E) = (e^{\beta E} + 1)^{-1}$ the Fermi function, 
$\xi({\mathbf k}) = {\mathbf k}^{2} /(2m) - \mu$, and $R_{pp}$ is
a remainder that converges to a finite value in the limit $k_{o} \rightarrow \infty$.
Note that $I_{pp} = - 1 + {\mathcal O}(k_{o}^{-1})$ owing to Eq.~(\ref{vo}). 

By the same token, for the particle-hole rung of Fig.1b we obtain:
\begin{eqnarray}
& &I_{ph}({\mathbf k_{1}}, \omega_{n_{1}}; {\mathbf k_{2}}, \omega_{n_{2}}) = 
\frac{1}{\beta} \sum_{\omega_{n}} \int \frac{d{\mathbf k}}{(2 \pi)^{3}}
\, v({\mathbf k})
\label{I-ph} \\ 
& &\times \, {\mathcal G}^{o}({\mathbf k_{1} + k}, \omega_{n_{1}} + 
\omega_{n}) \,
  {\mathcal G}^{o}({\mathbf k_{2} + k}, \omega_{n_{2}} + \omega_{n})        \nonumber \\
& &=   v_{o}\int^{k_{o}} \hspace{-0.3truecm}\frac{d{\mathbf k}}{(2 \pi)^{3}} \,
\frac{ \left[ f_{F}(\xi({\mathbf k_{1} + k})) \, - \, f_{F}(\xi({\mathbf k_{2} + k})) \right]}
{i(\omega_{n_{2}} \, - \, \omega_{n_{1}}) \, + \, \xi({\mathbf k_{1} + k}) \, - \,
\xi({\mathbf k_{2} + k})}\nonumber 
\end{eqnarray}

\noindent
which vanishes in the limit $k_{o} \, \rightarrow \, \infty$  owing to the vanishing 
of the prefactor $v_{o}$. 
In this case, the integral over ${\mathbf k}$ on the right-hand side of Eq.~(\ref{I-ph})
remains finite in the limit, because $f_{F}(\xi({\mathbf k}))$ is proportional to 
$\exp~{\{- \beta {\mathbf k}^{2}/(2m)\}}$ for large $|{\mathbf k}|$ (irrespective of 
the value of the chemical potential).
This situation has to be contrasted with the original argument by Galitskii,
whereby for a finite-range potential the particle-hole rung $I_{ph}$ does not vanish,
but it is reduced with respect to its particle-particle counterpart $I_{pp}$ 
by a factor $k_{F} a_{F}$ ($k_{F}$ being the Fermi wave vector).\cite{Galitskii}
In the weak-coupling limit, in fact, the integral on the right-hand side of
Eq.~(\ref{I-ph}) is proportional to $k_{F}$ from dimensional considerations. 

In an analogous way, one can show that in the particle-particle channel the 
contributions of the vertex corrections (like the one depicted in Fig.~\ref{fig:rung}c)
and of the two-particle effective interactions other than the rung (like the one
depicted in Fig.~\ref{fig:rung}d) vanish for our choice of the potential, since they 
both contain a factor $I_{ph}$ of the particle-hole type. [Note that the diagram of 
Fig.~\ref{fig:rung}c contains also a forbidden interaction between parallel spins.]

\begin{figure}
\narrowtext
\epsfxsize=3.3in 
\epsfbox{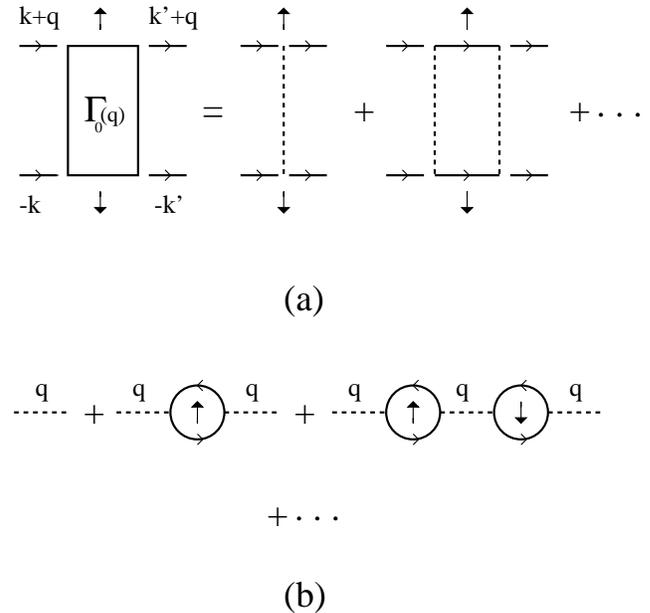}
\vspace{.2truecm}
\caption{\emph{(a)} Particle-particle ladder and \emph{(b)} 
series of particle-hole bubbles. Note that for a point-contact potential the 
particle-particle ladder depends only on the sum of the incoming (outgoing) 
four-momenta. Conventions are as in Fig.~\ref{fig:rung}.}
\label{fig:pplad}
\end{figure}

Despite these drastic simplifications, care should be exerted in not dismissing too 
promptly contributions which, albeit being individually vanishing as 
$k_{o} \, \rightarrow \, \infty$, belong to an infinite sequence that could  
instead converge to a finite value. 
Two examples of alternative behaviors are given in Fig.~\ref{fig:pplad}a and 
Fig.~\ref{fig:pplad}b. In particular, for the particle-particle ladder of 
Fig.~\ref{fig:pplad}a we obtain from Eq.~(\ref{I-pp}) (apart from an overall 
minus sign originating from the fermionic diagrammatic rules) :

\begin{eqnarray}
& &\frac{v_{o}}{1 \, + \, I_{pp}(k_{1}, k_{2})}\nonumber\\
& &= - \, 
\frac{\frac{2 \pi^{2}}{m k_{o}} \, + \, \frac{\pi^{3}}{m a_{F} k_{o}^{2}}}
{1 \, - \, \left( \frac{2 \pi^{2}}{m k_{o}} \, + \, \frac{\pi^{3}}{m a_{F} k_{o}^{2}} 
\right) \left( \frac{m k_{o}}{2 \pi^{2}} + R_{pp}(k_{1} + k_{2}) \right)}  \nonumber \\
& &= \, \frac{1}{\frac{m}{4 \pi a_{F}} \, + \, R_{pp}(k_{1} + k_{2})}  
                                                                  \label{p-p ladder}
\end{eqnarray}

\noindent
since both numerator and denominator in Eq.~(\ref{p-p ladder}) vanish like 
$k_{o}^{-1}$ when $k_{o} \, \rightarrow \, \infty$. 
[Note that we have introduced the four-vector notation 
$k \equiv ({\mathbf k}, \omega_{n})$. Note also that we have attached the $\uparrow$
spin to the upper line and the $\downarrow$ spin to the lower line of the ladder,
respectively. This convention will be consistently mantained in the following.]
For the particle-hole bubble series of Fig.~\ref{fig:pplad}b we obtain instead 
a vanishing result, since

\begin{equation}
\frac{v_{o}}{1 \, + \, I_{ph}(q;0)} \, \longrightarrow \, 0        \label{p-h bubble}
\end{equation}

\noindent
as $k_{o} \, \rightarrow \, \infty$.\cite{footnote-4}

The particle-particle ladder (\ref{p-p ladder}), upon surviving the limit
$k_{o} \, \rightarrow \, \infty$, represents a building block of the 
many-body diagrammatic structure. 
It is thus relevant to examine in detail its analytic behavior in the 
weak- and strong-coupling limits, in the order.
Let 

\begin{equation}
E_{\alpha} \, = \, i(\omega_{n_{1}} \, + \, \omega_{n_{2}}) \, - \, 
\frac{({\mathbf k_{1}} + {\mathbf k_{2}})^{2}}{4 m} \, + \, 2\mu      \label{E-alpha}
\end{equation}

\noindent
be the (complex) energy entering the denominator in Eq.~(\ref{I-pp}) and 
$k_{\alpha} = \sqrt{2 m E_{\alpha}}$ the associated (complex) wave vector.
In the weak-coupling limit,\cite{footnote-mu} both the $f_{F}$ terms and the $-1$ 
term within brackets in Eq.~(\ref{I-pp}) contribute to $R_{pp}$, yielding contributions 
proportional to $k_{F}$ and $k_{\alpha}$, respectively. 
One then obtains the following \emph{finite result\/} for the 
particle-particle ladder (\ref{p-p ladder}) 

\begin{eqnarray}
& &\frac{1}{\frac{m}{4 \pi a_{F}} + R_{pp}(k_{1} + k_{2})} 
\nonumber\\
& &\cong \frac{4 \pi a_{F}}{m} \left( \phantom{\frac{1}{1}}
                1 + {\mathcal O}(k_{F} a_{F}) +  
                {\mathcal O}(k_{\alpha} a_{F}) \right) \, , \label{pp-wc}
\end{eqnarray}                
 
\noindent
which permits one to classify the diagrammatic structure in powers of $k_{F} a_{F}$ 
(and/or of $k_{\alpha} a_{F}$) in the weak-coupling limit.\cite{Galitskii}

In the strong-coupling limit (whereby $\beta \mu \rightarrow - \, 
\infty$),\cite{footnote-mu} on the other hand, only the $-1$ term within brackets 
in Eq.~(\ref{I-pp}) contributes to $R_{pp}$, yielding

\begin{equation}
R_{pp}(k_{1} + k_{2}) \, = \, i \, \frac{m^{3/2}}{4 \pi} \,  
             \sqrt{E_{\alpha}} \, \mathrm{sgn}(\mathrm{Im}E_{\alpha})              \label{R-pp-sc}
\end{equation}

\noindent
with the standard convention for the cut of the square root along the negative
real axis.
Recalling further that $2 |\mu| = (m a_{F}^{2})^{-1}$ in the (extreme) strong-coupling
limit,\cite{PS-96} we obtain for the particle-particle ladder (\ref{p-p ladder}) 
the following \emph{polar structure\/}

\begin{equation}
\frac{1}{\frac{m}{4 \pi a_{F}}  +  R_{pp}(k_{1} + k_{2})} \cong  
\frac{8 \pi}{m^{2} a_{F}}  \frac{1}{i(\omega_{n_{1}} + \omega_{n_{2}}) -  
\frac{({\mathbf k}_{1} + {\mathbf k}_{2})^{2}}{4 m}}               \label{pp-sc}
\end{equation}

\noindent 
which (apart from the residue being different from unity)
resembles a ``free'' boson propagator with Matsubara frequency
$\Omega_{\nu} = \omega_{n_{1}} + \omega_{n_{2}} = 2\nu \pi \beta^{-1}$ ($\nu$ integer), 
wave vector ${\mathbf q} = {\mathbf k_{1}} + {\mathbf k_{2}}$, mass $2 m$, and vanishing
chemical potential.
[More generally, the bosonic chemical potential $\mu_{B} = 2 \mu + \epsilon_{o}$
should be added to the denominator of Eq.~(\ref{pp-sc}), where $\epsilon_{o}$ is the
bound-state energy of the two-fermion problem - cf. subsection IIIB. 
In the extreme strong-coupling limit $\mu_{B}$ vanishes.]
The complete analogy between the fermionic generalized particle-particle ladder 
(of which the ``bare'' particle-particle ladder (\ref{p-p ladder}) is only a piece) 
and the ``full'' boson propagator is discussed in Appendix A.
                                 
It is evident from the above considerations that the classification of diagrams
developed for a ``dilute'' system in weak coupling (which relies on the finite value 
(\ref{pp-wc}) of the particle-particle ladder - cf. Ref.\ 6) can no longer
be utilized in the strong-coupling limit, since in this case the particle-particle 
ladder does not reduce to a constant but develops a polar structure [cf. Eq.~(\ref{pp-sc})].
\emph{In the strong-coupling limit, a different classification scheme is therefore 
required to organize the diagrammatic structure for a ``dilute'' system\/}.   

\begin{figure}
\narrowtext
\epsfysize=4.0in\hspace{0.5in} 
\epsfbox{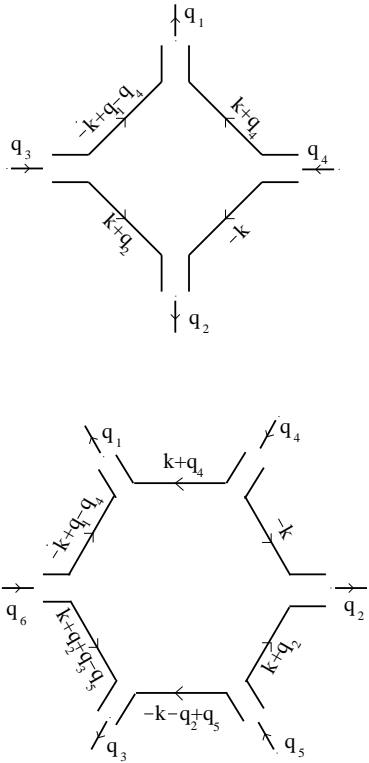}
\vspace{.2truecm}
\caption{Four- and six-point vertices for composite bosons. 
Incoming and outgoing bosonic four-momenta are indicated, 
as well as four-momenta on each fermionic line. 
Spin labels are understood to alternate on successive fermionic lines.}
\label{fig:Hikami}
\end{figure}

It is also evident from the above considerations that, with our choice of the 
fermionic interaction, the \emph{skeleton structure\/} of the diagrammatic theory can 
be constructed only with the particle-particle ladder (\ref{p-p ladder}) plus an 
infinite number of interaction vertices, like the ones depicted in 
Fig.~\ref{fig:Hikami} (besides \emph{one} spare single-particle fermionic Green's 
function that enters the fermionic self-energy diagrams, which in turn contains in 
principle all self-energy insertions originating from self-consistency).

The ladder and the vertices of Fig.~\ref{fig:Hikami} contain, by construction, only 
``bare'' single-particle fermionic Green's functions.
These vertices serve to connect the ladders among themselves, thus generating
complex diagrammatic structures.
In analogy with the so-called Hikami vertices occurring in the weak-localization 
problem,\cite{Hikami} we refer to these vertices as the four, six, ..., -point vertices, 
in the order.

As far as the dressing of the particle-particle ladder is concerned, the above 
statements can be justified in a rigorous fashion by resorting, e.g., to
functional-integral methods. 
To this end, in the following subsection we recall some relevant results 
obtained in Ref.\ 16, adapting them to the purposes of the present paper.
\subsection{Mapping onto a bosonic system via functional integrals}

We briefly review the procedure of Ref.\ 16 to extract the set of effective 
bosonic interactions mentioned above from the original fermionic action.
We first note that comparison with the results of Ref.\ 16 requires us to 
interpret the Fourier transform of the effective attractive potential in 
Eq.~(\ref{Eq:Hamiltonian}) as a ``separable'' potential, of the form

\begin{equation}  
\frac{1}{{\mathcal V}} \, V_{{\mathrm eff}}({\mathbf k} - {\mathbf k'}) \, \longrightarrow \, 
                    V \, w({\mathbf k}) \, w({\mathbf k'})         \label{sep-potential}                                          
\end{equation}

\noindent
where ${\mathcal V}$ is the quantization volume, $V$ is a negative constant, and

\[	w({\mathbf k}) \, = \, \left\{ \begin{array}{ll}
                        1  &  \mbox{if $|{\mathbf k}| < k_{o}$ \, ,}   \\
                        0  &  \mbox{otherwise \, .}
                                  \end{array} \right.            \]

\noindent
Note that the limit $k_o \rightarrow \infty$ recovers our ``contact'' potential, 
provided we identify $V = v_{o} / {\mathcal V}$, with $v_{o}$ given by (\ref{vo}).

In Ref.\ 16 the original fermionic partition function was written in the form 
of a functional integral

\begin{equation}
{\mathcal Z} \, = \, \int {\mathcal D}\bar{c} {\mathcal D}c \,\, \exp\{-S\}
                                                \label{partition-function}  
\end{equation}

\noindent
with action

\begin{equation}	
S \, = \, \int_0^\beta d\tau  \left( \sum_{{\mathbf k},\sigma} 
                         \bar{c}_{\sigma}({\mathbf k},\tau) 
	\frac{\partial}{\partial \tau} c_{\sigma}({\mathbf k},\tau) + H(\tau) \right) \, ,
	                                             \label{action}  
\end{equation}
	
\noindent
where $\tau$ is the imaginary time and $(c,\bar{c})$ are Grassmann variables.
Owing to the form (\ref{sep-potential}) of the interaction potential, the quartic part 
of the Hamiltonian was further expressed in terms of the operator

\begin{equation} 
{\mathcal B}({\mathbf q},\tau) \, = \, \sum_{{\mathbf k}} w({\mathbf k}) 
c_{\downarrow}(-{\mathbf k} + {\mathbf q}/2, \tau)
c_{\uparrow}({\mathbf k} + {\mathbf q}/2, \tau)  \,\,\, .   \label{quadratic-operator}  
\end{equation}

\noindent
With the use of the following Hubbard-Stratonovich transformation (in the 
particle-particle channel)

\begin{eqnarray}
& &\exp\left\{- V \bar{{\mathcal B}}({\mathbf q},\tau) {\mathcal B}({\mathbf q},\tau) \right\}
=  - \frac{1}{\pi V} \int d b^*({\mathbf q},\tau) d b({\mathbf q},\tau) 
\label{Hub-Straton} \\
& &\times\exp\left\{\frac{1}{V} |b({\mathbf q},\tau)|^2 + b({\mathbf q},\tau) 
\bar{{\mathcal B}}({\mathbf q}, \tau)  +  
b^*({\mathbf q},\tau) {\mathcal B}({\mathbf q},\tau) \right\}     \nonumber 
\end{eqnarray}

\noindent
where $b({\mathbf q},\tau + \beta) = b({\mathbf q},\tau)$ for any ${\mathbf q}$ and 
$\tau$, the Grassmann variables were eventually integrated out, to obtain (apart
from irrelevant numerical constants)

\begin{equation}  
{\mathcal Z} \, = \, \int {\mathcal D} b^* {\mathcal D} b \,\, \exp\{ - S_{\mathrm{eff}} \}
                                                 \label{eff-partition-function}     
\end{equation}

\noindent
with the ``effective'' bosonic action

\begin{equation}
S_{\mathrm{eff}} \, = \, - \frac{1}{\beta V} \sum_q |b(q)|^2 - \mathrm{tr} \ln {\mathbf M} \, . 
                                                         \label{eff-action} 
\end{equation}

\noindent 
In Eq.~(\ref{eff-action}), we have set $q \equiv ({\mathbf q},\Omega_\nu)$, the matrix 
${\mathbf M}$ is given by

\begin{equation}      
{\mathbf M}(k,k') \, = \, \left( \begin{array}{ccc}
\epsilon(k)\delta_{k,k'} & , &
\frac{1}{\beta} w(\frac{k+k'}{2}) b^* (k-k') \\
\frac{1}{\beta} w(\frac{k+k'}{2}) b (k'-k) & , &
-\epsilon(-k)\delta_{k,k'} 
\end{array} \right)                                 \label{Matrix}                         
\end{equation}

\noindent
with $\epsilon(k) = i \omega_{n} -\xi({\mathbf k})$, and the trace is over the 
four-momentum and the matrix indices. 
The {\it effective\/} action $S_{\mathrm{eff}}$ contains \emph{all powers\/} in 
the bosonic-like variables $b$. 
Expanding the logarithm in Eq.~(\ref{eff-action}) in fact yields

\begin{equation}  
S_{\mathrm{eff}} \, = \, - {\rm tr} \ln {\mathbf M}_S \, - \,  \frac{1}{\beta V} \sum_q |b(q)|^2
\, + \, \sum_{l=1}^\infty \frac{1}{2l} {\rm tr} X^{2l}    \label{expansion-eff-action}
\end{equation}

\noindent
where

\begin{equation}  
{\mathbf M}_S (k,k') \, = \,
\left(  \begin{array}{cc}
\epsilon(k) & 0 \\
0 & -\epsilon(-k)
\end{array}  \right)  \delta_{k,k'}               \label{diagonal-matrix} 
\end{equation}

\noindent
and

\begin{equation} 
X(k,k') = w\left( \frac{k+k'}{2} \right)
\left(  \begin{array}{cc}
0 & \frac{b^*(k-k')}{\beta \epsilon(k)} \\
- \frac{b(k'-k)}{\beta \epsilon(-k)} & 0
\end{array}    \right)    \,\,\, .                \label{off-diagonal-matrix} 
\end{equation}

\begin{figure}
\narrowtext
\epsfxsize=3.3in 
\epsfbox{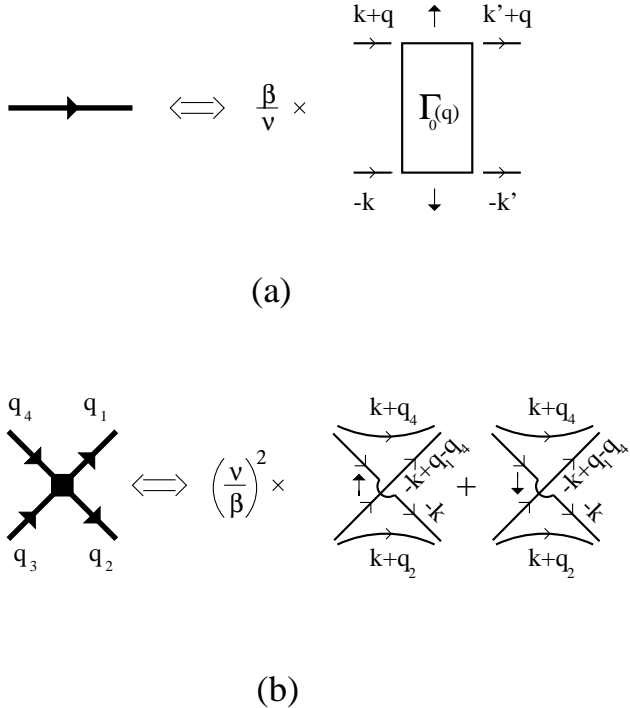}
\vspace{.2truecm}
\caption{ Graphical correspondence \emph{(a)} between the 
``bare'' propagator for composite bosons (represented by a thick line) and the
fermionic particle-particle ladder of Fig.~\ref{fig:pplad}a, and \emph{(b)}
between the effective 
two-boson interaction and the four-point vertex of Fig.~\ref{fig:Hikami},
where now the spin labels have been explicitly indicated in the internal lines.
Note that the fermionic lines composing the four-point vertex have been 
rearranged with respect to Fig.~\ref{fig:Hikami}, in order to resemble the 
bosonic vertex more closely
(accordingly, the fermionic lines never intersect each other in the four-point
vertex).
Appropriate powers of $\beta/{\mathcal V}$ as required by Eqs.~(\ref{bare bb})
 and
(\ref{two-body-potential}) have been indicated explicitly.}         
\label{fig:cobos}
\end{figure}

Consider the \textit{quadratic\/} terms first. The ensuing quadratic action reads:

\begin{equation}  
S^{(2)}_{\mathrm{eff}} \, = \, \frac{1}{\beta^2} \sum_q |b(q)|^2 \, A(q) 
                                                     \label{quadratic-eff-action} 
\end{equation}

\noindent
where

\begin{eqnarray}
A(q)&=&-\frac{\beta}{V}\nonumber\\
&-&\beta\sum_{{\mathbf k}} w^2({\mathbf k}-{\mathbf q}/2)
	   \frac{\left[f_{F}(\xi({\mathbf k}))+
f_{F}(\xi({\mathbf k}-{\mathbf q}))-1\right]}	
{i\Omega_{\nu}-\xi({\mathbf k})-\xi({\mathbf k}-{\mathbf q})}\nonumber\\ 
& = & - \frac{\beta}{V} \left(\phantom{\frac{1}{1}}\hspace{-0.2truecm} 
                                     1 \,+ \, I_{pp}(q) \right) \label{A}
\end{eqnarray}

\noindent
with $I_{pp}$ given by Eq.~(\ref{I-pp}). The ``bare'' bosonic propagator is thus
expressed in terms of the particle-particle ladder (\ref{p-p ladder}), by writing

\begin{equation}
<b^*(q) b(q)>_{S^{(2)}_{\mathrm{eff}}} \, = \, - \, \frac{\beta}{{\mathcal V}} \, 
\frac{v_{o}}{1 \, + \, I_{pp}(q)} \, \equiv \, \frac{\beta}{{\mathcal V}} 
                       \,\,  \Gamma_{o}(q)                         \label{bare bb}
\end{equation}

\noindent
where $\Gamma_{o}(q)$ is the particle-particle ladder depicted in 
Fig.~\ref{fig:pplad}a 
(including now the appropriate overall sign that was missing in 
Eq.~(\ref{p-p ladder})). 
This identification is shown graphically in Fig.~\ref{fig:cobos}a for a 
definite choice of the spin labels (this convention will be mantained in the 
rest of the paper).
Note that the factor $\beta/{\mathcal V}$ of Eq.~(\ref{bare bb}) enters also the 
definition of the ``full'' bosonic propagator discussed in Appendix A.

It is important to emphasize that the identification (\ref{bare bb}) (as well as 
the other results of this subsection and of Appendix A) holds \emph{irrespective\/} 
of the value of the fermionic scattering length $a_{F}$. 
Nonetheless, referring to a system of interacting composite bosons acquires 
physical meaning in the strong-coupling limit only.

For the \emph{quartic\/} terms we obtain instead from Eq.~(\ref{expansion-eff-action})

\begin{eqnarray}
S_{\mathrm{eff}}^{(4)}&=& \frac{1}{4} {\rm tr} X^4 = \frac{1}{4 \beta 
{\mathcal V}} \sum_{q_{1} \dots q_{4}}\nonumber\\
&\times&      
\tilde{u}_{2}(q_{1} \dots q_{4}) b^*(q_1) b^*(q_2) b(q_3) b(q_4)         
\label{quartic-eff-action}
\end{eqnarray}
	
\noindent
where the (four-point) \emph{effective two-boson interaction\/} is given by
[cf. Fig.~\ref{fig:Hikami}]

\begin{eqnarray}   
& &\tilde{u}_{2}(q_{1} \dots q_{4}) \, = \, \delta_{q_1+q_2,q_3+q_4} 
\left(\frac{{\mathcal V}}{\beta}\right)^{2}  \frac{2}{\beta {\mathcal V}}
\nonumber\\
& &\times \sum_k 
\frac{1}{\epsilon(-k)\epsilon(k+q_2)\epsilon(-k+q_1-q_4)\epsilon(k+q_4)}
\label{two-body-potential}     
\end{eqnarray}

\noindent
with the limit $k_{o} \rightarrow \infty$ already taken in the last expression.
It is worth noting the following features of the above expressions: 
\emph{(i)} The factor $1/4$ (instead of $1/2$) in Eq.~(\ref{quartic-eff-action}) 
applies to a \emph{symmetrized\/} bosonic two-body interaction, which makes the 
counting of the bosonic Feynman diagrams easier \cite{Popov-1,Popov-2};
\emph{(ii)} The interaction (\ref{two-body-potential}) is indeed symmetric under
the interchange $q_{1} \leftrightarrow q_{2}$ or $q_{3} \leftrightarrow q_{4}$, as
it can be explicitly verified; 
\emph{(iii)} The interaction (\ref{two-body-potential}) depends on wave vectors 
\emph{as well as\/} on Matsubara frequencies, revealing in this way the composite 
nature of the bosons; 
\emph{(iv)} The factors $\beta$ and ${\mathcal V}$ entering 
Eqs.~(\ref{quartic-eff-action}) and (\ref{two-body-potential}) are
needed to comply with the standard bosonic and fermionic diagrammatic rules and
with the definition of the ``full'' bosonic propagator given in Appendix A;
\emph{(v)} The energy denominators in Eq.~(\ref{two-body-potential}) correspond to 
single-particle (bare) fermionic Green's functions, since 
${\mathcal G}^{o}(k) = \epsilon(k)^{-1}$;
\emph{(vi)} The factor of $2$ in the definition (\ref{two-body-potential}) corresponds
to the two different sequences of spin labels that can be attached to the four
fermionic Green's functions, as shown graphically in Fig.~\ref{fig:cobos}b 
(where the identification with the effective two-boson interaction is also indicated). 
Keeping track of the spin labels will, in fact, prove important in the following to 
establish the desired mapping between the bosonic and fermionic diagrammatic 
structures.
When considering the sum $S_{\mathrm{eff}}^{(2)}+S_{\mathrm{eff}}^{(4)}$ 
of the quadratic and quartic actions, the bosonic propagator 
$<b^*(q) b(q)>_{S^{(2)}_{\mathrm{eff}} + S^{(4)}_{\mathrm{eff}}}$ can 
be expressed in terms of the ``bare'' bosonic propagator (\ref{bare bb}) and of the 
effective two-boson interaction (\ref{two-body-potential}) via Wick's theorem. 
The topology of the resulting diagrammatic structure, as well as the symmetry 
factor of each diagram, are identical to those obtained for true (point-like)
bosons.\cite{Popov-1,Popov-2}
The associated fermionic diagrammatic structure can then be constructed 
whenever needed by the correspondence rules shown in Fig.~\ref{fig:cobos}.

\begin{figure}
\narrowtext
\epsfysize=4.0in\hspace{0.32in}
\epsfbox{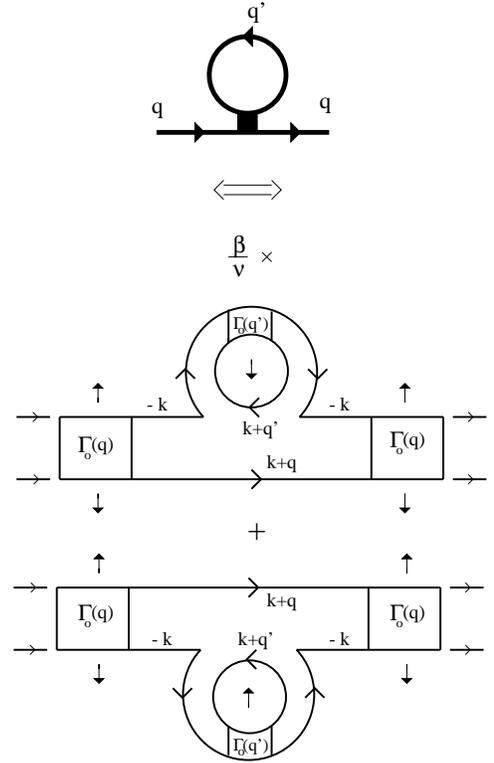}
\vspace{.2truecm}
\caption{Graphical correspondence between the composite-boson 
propagator and the two-fermion Green's function in the particle-particle channel, to 
first order in the \emph{four-point\/} interaction vertex.
This vertex can be identified from Fig.~\ref{fig:Hikami} by setting $q_2=q_3=q$ and
$q_1=q_4=q'$ therein.}         
\label{fig:bosHF}
\end{figure}

As an example of this correspondence, we show in Fig.~\ref{fig:bosHF} the 
bosonic propagator to first order in the interaction, together with the 
associated fermionic 
diagrams for the two-fermion Green's function in the particle-particle channel.
Since the symmetry factor of this bosonic diagram is unity, only two fermionic 
diagrams are associated with it. 
[The value of the bosonic symmetry factor and the number of independent fermionic 
diagrams are intimately connected, as it will be discussed in Appendix B.]
Note that a factor $1/(\beta {\mathcal V})$ is attached to the sum over $q'$ in both
bosonic and fermionic diagrams.
Note also that the minus sign, which is associated with one power of the interaction
in the bosonic diagram, is associated instead with the presence of a closed loop in 
the corresponding fermionic diagrams (the minus signs associated with the fermionic
interaction being already taken into account in the definition of the particle-particle 
ladder).
Note finally that the bosonic self-energy insertion of Fig.~\ref{fig:bosHF} has 
the same topological structure of the bosonic Hartree-Fock self-energy 
diagram.\cite{Popov-1,Popov-2}

A typical value of the two-boson effective interaction is obtained by considering
the strong-coupling limit $\beta \mu \rightarrow - \infty$ and setting
$q_1 = \cdots = q_4 = 0$ in Eq.~(\ref{two-body-potential}). One gets:

\begin{equation}
\tilde{u}_{2}(0) \, = \, 2 \, \left( \frac{{\mathcal V}}{\beta} \right)^{2}
    \left( \frac{m^{2} a_{F}}{8 \pi} \right)^{2} \, u_{2}(0)  \label{u2-tilde-limit}
\end{equation}

\noindent
where \cite{PS-96}

\begin{equation}  
u_{2}(0) \, = \, \frac{4 \pi (2a_{F})}{2m} \, .                \label{u2-limit}
\end{equation}

\noindent
The factor $m^{2} a_{F}/(8 \pi)$ in Eq.~(\ref{u2-tilde-limit}) reflects the difference 
between the true bosonic propagator and the particle-particle ladder in the 
strong-coupling limit [cf. also Eq.~(\ref{pp-sc})]. 
Owing to this difference, $u_{2}(0)$ given by Eq.~(\ref{u2-limit}) (and not
$\tilde{u}_{2}(0)$ given by Eq.~(\ref{u2-tilde-limit})) has to be 
identified with the boson-boson interaction at zero four-momenta.
We shall return to the difference between $\tilde{u}_{2}$ and $u_{2}$ in 
subsection IIIB.\cite{footnote-stability}

Recalling further that the scattering length $a^{\mathrm{Born}}_{B}$ within the Born 
approximation, obtained for a pair of true bosons (each of mass $2m$) mutually 
interacting via a two-body potential with Fourier transform $u_{2}(0)$ at zero wave 
vector, is given by $a^{\mathrm{Born}}_{B} = 2m u_{2}(0)/(4 \pi)$, Eq.~(\ref{u2-limit})
yields the following relation between the bosonic and fermionic scattering lengths:

\begin{equation}
a^{\mathrm{Born}}_{B} \, = \, 2 \, a_{F} \, .                             \label{a-Born}
\end{equation}

\noindent
The result (\ref{a-Born}) was also obtained in Ref.\ 9 within the
fermionic self-consistent T-matrix approximation (which corresponds to the bosonic 
Hartree-Fock approximation of Fig.~\ref{fig:bosHF} in the strong-coupling
limit - see the next subsection), where it was regarded to be the value of the 
scattering length $a_{B}$ for a ``dilute'' system of composite bosons. 
We will show in Section IV that the result (\ref{a-Born}) actually differs from 
$a_{B}$, when \emph{all\/} diagrams associated with a ``dilute'' system of 
composite bosons are taken into account.

Beside the four-point vertex (\ref{two-body-potential}), the composite nature of 
the bosons produces (an infinite set of) additional vertices.
Let us consider, in particular, the six-point vertex defined by keeping the terms
with $l = 3$ in Eq.~(\ref{expansion-eff-action}). We obtain:

\begin{eqnarray}
& &S_{\mathrm{eff}}^{(6)} = \frac{1}{6} {\rm tr} X^6 =
\frac{1}{6 (\beta {\mathcal V})^{2}} \sum_{q_{1} \dots q_{6}}\nonumber\\
& &\times
                                     \label{6-eff-action} 
\tilde{u}_{3}(q_{1} \dots q_{6}) b^*(q_1) b^*(q_2) b^*(q_3) b(q_4) b(q_5) b(q_6)         
\end{eqnarray}

\noindent
where the (six-point) \emph{effective three-boson interaction\/} is given by
[cf. Fig.~\ref{fig:Hikami}]
\begin{eqnarray}
& &\tilde{u}_{3}(q_{1} \dots q_{6}) = \delta_{q_1+q_2+q_3,q_4+q_5+q_6}
\left(\frac{{\mathcal V}}{\beta}\right)^{3}  \frac{2}{\beta {\mathcal V}} 
\sum_k\nonumber \\
& &\times \frac{(-1)}{\epsilon(-k)\epsilon(k+q_2)\epsilon(-k-q_2+q_5)}
\nonumber\\
& &\times\frac{1}{\epsilon(k+q_2+q_3-q_5)\epsilon(-k+q_1-q_4)\epsilon(k+q_4)} 
\, .  \label{three-body-potential}       
\end{eqnarray}

\begin{figure}
\narrowtext
\epsfysize=4.0in\hspace{0.125in} 
\epsfbox{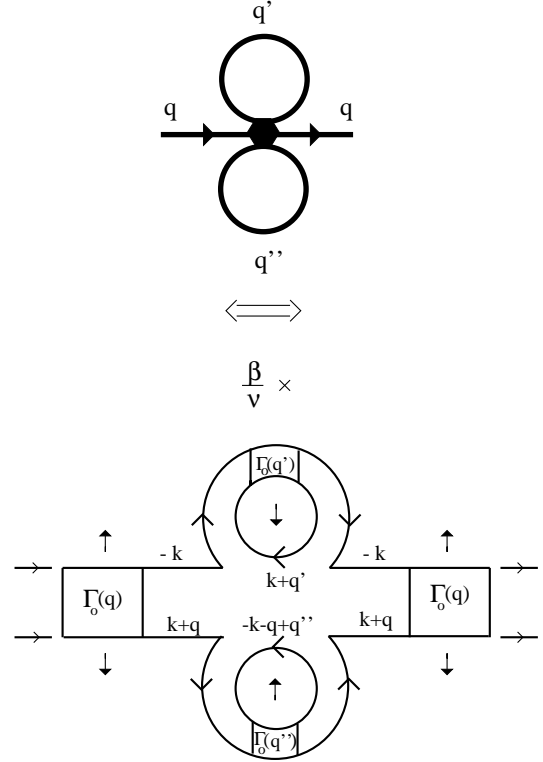}
\vspace{.2truecm}
\caption{Graphical correspondence between the 
composite-boson propagator and the two-fermion Green's function in the 
particle-particle channel, to first order in the \emph{six-point\/} interaction vertex.
This vertex can be identified from Fig.~\ref{fig:Hikami} by setting $q_2=q_6=q$,
$q_1=q_4=q'$, and $q_3=q_5=q''$ therein.} 
\label{fig:bosbutt}
\end{figure}

\noindent
In the strong-coupling limit, whereby $|\mu|$ is the relevant energy scale in the 
problem, from dimensional considerations we get 
$\tilde{u}_{2} (\beta/{\mathcal V})^{2} \sim |\mu|^{-3/2}$ and
$\tilde{u}_{3} (\beta/{\mathcal V})^{3} \sim |\mu|^{-7/2}$ (in three dimensions).
For this reason, the contribution of the tree-boson vertex (\ref{three-body-potential})
is suppressed with respect to the contribution of the two-boson vertex
(\ref{two-body-potential}).
To be more precise, one should compare the values of similar diagrams constructed 
with the four- and six-point vertices, respectively (like, for instance, the ones 
depicted in Fig.~\ref{fig:bosHF} and Fig.~\ref{fig:bosbutt}). 
The value of the diagram of Fig.~\ref{fig:bosbutt} is smaller than the
value of the diagram of Fig.~\ref{fig:bosHF} by the quantity

\begin{equation}
\frac{|\mu|^{-7/2} \, \epsilon_{F}^{3} \, |\mu|}
     {|\mu|^{-3/2} \, \epsilon_{F}^{3/2} \, |\mu|^{1/2}} \,\, \sim \,\, 
                                      (k_{F} a_{F})^{3}  \, .      \label{ratio 2-3}
\end{equation}

\noindent
Here, the factors containing the Fermi energy $\epsilon_{F}$ [$= k_{F}^{2}/(2m)
=(3 \pi^2 \rho)^{2/3}/(2m)$] originate from the bosonic cycles (cf. Section III), 
while the factors $|\mu|$ and $|\mu|^{1/2}$ originate from the residue in 
Eq.~(\ref{pp-sc}).
The diagram of Fig.~\ref{fig:bosbutt} can thus be neglected in comparison
to the diagram of Fig.~\ref{fig:bosHF}, since $k_{F}a_{F} \ll 1$ in the
strong-coupling limit.

The above argument can be made more general, by showing that \emph{all interaction 
vertices can be neglected in comparison with the four-point vertex in the 
strong-coupling limit\/}.\cite{footnote-5} 
In this limit, one can thus construct all diagrams representing the two-particle Green's
function in the particle-particle channel in terms of the ``bare'' ladder 
and of the four- point interaction vertex only.
This is precisely what one would expect on physical grounds, since the interactions 
involving more than two bodies become progressively less effective as the composite 
bosons overlap less when approaching the strong-coupling limit.
However, it is also clear from the above considerations that in the 
\emph{extreme\/} strong-coupling limit (whereby $a_{F} \rightarrow 0$) the four-point 
vertex can be neglected, too, and the composite bosons become 
\emph{effectively free\/}.\cite{footnote-1}

In the next subsection we will show how the (self-consistent) fermionic T-matrix 
approximation can be examined in terms of the four, six, $\cdots $, -vertex functions 
in the strong-coupling limit.
\subsection{Fermionic T-matrix approximation in the strong-coupling limit}

\begin{figure}
\narrowtext
\epsfysize=4.0in \hspace{0.125in}
\epsfbox{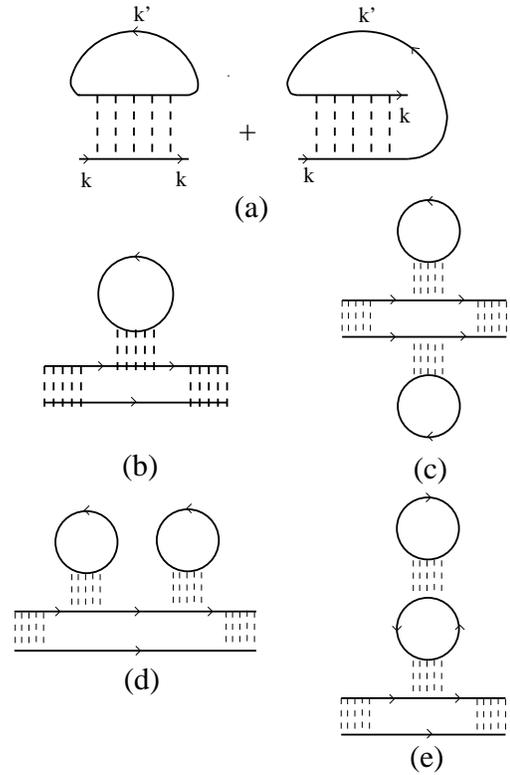}
\vspace{.2truecm}
\caption{\emph{(a)} Self-energy diagrams corresponding to the 
self-consistent fermionic T-matrix approximation (full lines here 
represent self-consistent fermionic single-particle Green's functions and
spin labels have been suppressed);
Self-energy corrections entering the particle-particle ladder, obtained by 
contracting the \emph{(b)} four-point vertex and \emph{(c-e)} six-point vertex
(full lines now represent the ``bare'' fermionic single-particle Green's
functions).}
\label{fig:ferTmat}
\end{figure}

The T-matrix approximation for a ``dilute'' Fermi gas represents one of 
the few cases in the many-body theory where the choice of the self-energy 
diagrams can be 
controlled by an external small parameter.\cite{FW}
In the original version by Galitskii,\cite{Galitskii} the T-matrix approximation was 
conceived for a \emph{repulsive\/} fermionic interaction of finite range (thus 
excluding bound states) and with the scattering length $a_{F}$ always positive 
(albeit small).
The fermionic self-energy diagram associated with this approximation is depicted 
at the left in Fig.~\ref{fig:ferTmat}a, and is obtained by closing
the particle-particle ladder with a single-particle Green's function in the
only possible way.
[The diagram at the right in Fig.~\ref{fig:ferTmat}a was included by Galitskii
original treatment,\cite{Galitskii} but vanishes for our choice of the attractive
potential since it contains forbidden interactions between parallel spins.
By the same token, no spin summation needs to be considered for the fermionic
loop at the left in Fig.~\ref{fig:ferTmat}a.]
In Fig.~\ref{fig:ferTmat}a all single-particle lines are regarded to be 
\emph{self-consistent\/}, and thus contain self-energy insertions of the same kind 
of the ones depicted in the figure.\cite{footnote-6}
In this way, the self-consistent fermionic T-matrix is ``conserving'' in the
Baym-Kadanoff sense.\cite{BK,Baym}
Recalling that (with our regularization of the potential) the particle-particle 
ladder depends only on the sum of the incoming (outgoing) four-momenta, the 
self-energy of Fig.~\ref{fig:ferTmat}a reads

\begin{equation}
\Sigma_{F}(k) \, = \, - \, \frac{1}{\beta {\mathcal V}} \, \sum_{k'} \,
\Gamma_{s}(k+k') \,\, {\mathcal G}(k')                        \label{Sigma-T-matrix}
\end{equation}

\noindent
where $\Gamma_{s}$ is obtained from $\Gamma_{o}$ by replacing everywhere the ``bare'' 
${\mathcal G}_{o}$ with the self-consistent ${\mathcal G}$.

In the approximation of the lowest order, self-energy insertions are neglected and 
the ``full'' single-particle lines of Fig.~\ref{fig:ferTmat}a are 
replaced by ``bare'' ones. 
The particle-particle ladder $\Gamma_{s}$ of Eq.~(\ref{Sigma-T-matrix})
thus reduces to $\Gamma_{o}$ of Fig.~\ref{fig:pplad}a, which (apart from a sign)
is approximately given by Eq.~(\ref{pp-wc}) in the weak-coupling limit
(irrespective of the sign of $a_{F}$). We thus obtain to this order:

\begin{equation}
\Sigma_{F}^{(o)}(k) \, = \, \frac{4 \pi a_{F}}{m} \, \frac{\rho}{2} \, = \,
\epsilon_{F} \left( \frac{4}{3 \pi} \, k_{F} a_{F} \, + \, 
{\mathcal O}((k_{F}a_{F})^{2}) \right)               \label{Sigma-T-matrix-F-0}
\end{equation}

\noindent
where the density factor $\rho$ [$=k_{F}^{3}/(3 \pi^{2})$] is supplied by the 
closed fermionic loop.
Considering self-energy insertions on top of the result (\ref{Sigma-T-matrix-F-0}) 
modifies $\Sigma_{F}/\epsilon_{F}$ only to ${\mathcal O}((k_{F}a_{F})^{2})$.

For the \emph{attractive\/} fermionic interaction of interest, the scattering 
length $a_{F}$ changes from being negative when the two-body problem fails to support
a bound state, to being positive when the bound state is eventually supported by
increasing the interaction strength.
As already discussed in subsection IIA, the sign change of $a_{F}$ has dramatic
effects on the ``bare'' particle-particle ladder $\Gamma_{o}$, that changes from the
behavior (\ref{pp-wc}) in the weak-coupling limit to developing a polar structure
in the strong-coupling limit (cf. Eq.~(\ref{pp-sc})).
In the latter limit, the self-energy $\Sigma_{F}^{(o)}$ acquires the density factor
from the particle-particle ladder and not from the closed fermionic loop.
This can be explicitly seen by transforming the Matsubara frequency sum in the 
expression for $\Sigma_{F}^{(o)}$ into a contour integration in the complex 
frequency plane as follows:

\begin{eqnarray}
& &\Sigma_{F}^{(o)}(k) = - \, \frac{1}{\beta {\mathcal V}} \, \sum_{k'} \, 
\Gamma_{o}(k+k') \,\,  {\mathcal G}^{o}(k')=-\! 
\int\frac{d{\mathbf k'}}{(2 \pi)^{3}}\nonumber\\
&\times &\oint_{C}\!
\frac{dz}{2 \pi i} \, \Gamma_{o}({\mathbf k}+{\mathbf k'},z) \, 
{\mathcal G}^{o}({\mathbf k'},z-i\omega_{n}) \,
\frac{1}{e^{\beta z} - 1}                               \label{Sigma-T-matrix-B-0}
\end{eqnarray}

\noindent
where $C$ is a contour encircling the imaginary axis in the positive sense.
In the strong-coupling limit, ${\mathcal G}^{o}({\mathbf k'},z-i\omega_{n})$
has a pole for Re($z$) $\ge |\mu|$, so that the contribution of this pole to
the integral in Eq.~(\ref{Sigma-T-matrix-B-0}) is strongly reduced by the Bose
factor $\exp\{-\beta|\mu|\}$. 
The dominant contribution to the integral in Eq.~(\ref{Sigma-T-matrix-B-0})
thus comes from the pole of $\Gamma_{o}({\mathbf k}+{\mathbf k'},z)$, yielding

\begin{eqnarray}
\Sigma_{F}^{(o)}(k) & \cong & \frac{8 \pi}{m^{2} a_{F}} \, 
\int\! \frac{d{\mathbf k'}}{(2 \pi)^{3}} \,\, 
\frac{1}{i\omega_{n}+\xi({\mathbf k'})-\frac{({\mathbf k}+{\mathbf k'})^{2}}{4m}}
\nonumber\\ 
&\times& \frac{1}{e^{\beta ({\mathbf k}+{\mathbf k'})^{2}/(4m)} - 1}      \nonumber \\
& \approx & \frac{8 \pi}{m^{2} a_{F}} \,\, \frac{1}{i\omega_{n}+\xi({\mathbf k})}
\,\, \frac{\rho}{2}   \,\, .                        \label{Sigma-T-matrix-B-0-approx}
\end{eqnarray}

It is interesting to note that the expression (\ref{Sigma-T-matrix-B-0-approx})
for the self-energy leads to a \emph{double-fraction structure\/} in the  
single-particle fermionic Green's function, since

\begin{equation}
{\mathcal G}({\mathbf k},\omega_{n}) \, = \, \frac{1}{i\omega_{n}-\xi({\mathbf k})-
\frac{B}{i\omega_{n}+\xi({\mathbf k})}}                  \label{G-double-fraction}
\end{equation}

\noindent
with $B=4 \pi \rho/(m^{2} a_{F})$. 
It is then clear that the expression (\ref{G-double-fraction}) has the form of the 
BCS Green's function, namely,

\begin{equation}
{\mathcal G}({\mathbf k},\omega_{n}) \, = \, 
\frac{u^{2}({\mathbf k})}{i\omega_{n}-E({\mathbf k})} \, + \,
\frac{v^{2}({\mathbf k})}{i\omega_{n}+E({\mathbf k})}   \,\, ,         \label{G-BCS}
\end{equation}

\noindent
where
\begin{equation}
E({\mathbf k}) \, = \, \sqrt{\xi^{2}({\mathbf k}) + B}            \label{E-BCS}
\end{equation}

\noindent
has a gap given approximately by $|\mu| + B/(2|\mu|)$ and where

\begin{equation}
v^{2}({\mathbf k}) \, = \, 1 - u^{2}({\mathbf k}) \, \cong \, 
\frac{4}{3 \pi} \, \frac{(k_{F}a_{F})^{3}}
{[1+(|{\mathbf k}| a_{F})^{2}]^{2}}  \, .                             \label{v2}
\end{equation}

\noindent
Besides the main peak at $E({\mathbf k})$ (with weight $u^{2}({\mathbf k})$), the 
spectral function associated with the Green's function (\ref{G-BCS}) has also 
a secondary peak at $-E({\mathbf k})$ (with weight $v^{2}({\mathbf k})$).
This secondary peak is relevant to photoemission experiments and can be interpreted
as a  ``shadow-band'' structure, associated with a real gap in the single-particle 
spectrum; this gap, in turn, might be regarded as anticipating the pseudo-gap actually
observed in ARPES experiments above the superconducting transition 
temperature.\cite{Ding,Loeser} 
It is also for this promising feature that several authors have considered the 
fermionic T-matrix approximation as representing the ``dilute-approach approximation'' 
for the BCS-BE crossover 
problem.\cite{Fresard,Haussmann,Micnas,Levin-97,Randeria-97-1,KKK} 

Self-energy insertions associated with the self-consistency of the fermionic 
single-particle Green's function modify both $\Gamma_{o}$ and ${\mathcal G}^{o}$ in
Eq.~(\ref{Sigma-T-matrix-B-0}).
As far as the self-energy insertions on $\Gamma_{o}$ are concerned, they can be
interpreted in terms of the four, six, $\cdots$, -point interaction vertices
discussed in the previous subsections. 
Typical examples are shown in Figs.~\ref{fig:ferTmat}b-e, where the
``bare'' single-particle lines associated with the vertices are marked by arrows.
Note, in particular, that the diagram of Fig.~\ref{fig:ferTmat}b
corresponds to a contraction of the four-point vertex, while the diagrams of
Figs.~\ref{fig:ferTmat}c-e correspond to all possible contractions
of the six-point vertex.
Additional diagrams not shown in the figure would then contain interaction vertices
of higher order.
Note also that the diagrams of Figs.~\ref{fig:ferTmat}b-c have been
already considered in Fig.~\ref{fig:bosHF} and in Fig.~\ref{fig:bosbutt},
respectively.
From the results of the previous subsection we then conclude that only the
diagram of Fig.~\ref{fig:ferTmat}b needs to be retained in the
strong-coupling limit, the other diagrams being suppressed with respect to it
at least by a factor $(k_{F} a_{F})^{3}$ (cf. Eq.~(\ref{ratio 2-3})).
The diagram of Fig.~\ref{fig:ferTmat}b corresponds to the Hartree-Fock 
approximation for the self-energy of composite bosons.\cite{footnote-2}
Note finally that self-energy insertions on ${\mathcal G}^{o}$ in
Eq.~(\ref{Sigma-T-matrix-B-0}) could be interpreted similarly, in an open-ended way.

The above argument leads us to the conclusion that the fermionic T-matrix approximation 
reproduces the Hartree-Fock approximation to the self-energy of composite bosons 
in the strong-coupling limit. 
This self-energy modifies the pole of $\Gamma_{o}$ in Eq.~(\ref{Sigma-T-matrix-B-0})
and affects eventually the (pseudo) gap structure in the fermionic self-energy.
There exist, however, \emph{additional\/} contributions to the self-energy of
composite bosons which are of the \emph{same order\/} (in the small parameter
$k_{F} a_{F}$) of the Hartree-Fock approximation just discussed.
These contributions are not included in the fermionic T-matrix approximation and
must be considered separately, as shown in the next Section.
\section{T-matrix approximation for composite bosons}

In this Section we shall set up an approximation for the fermionic self-energy, which
describes the ``low-density'' regime \emph{both} in the weak-coupling (fermionic) and in
the strong-coupling (bosonic) limits on equal footing.
To this end, we rely on the results of the previous Section and consider first the 
construction of the two-fermion Green's function in the bosonic limit, in terms of 
the ``bare'' particle-particle ladder and of the effective two-boson interaction.
Before examining this procedure in detail, however, it is instructive to briefly 
recall some well-known results concerning the self-energy for a ``dilute'' system 
of true (point-like) bosons,\cite{Beliaev,Popov-1,Popov-2} and in particular the
method for selecting the relevant self-energy diagrams for this system.
In this way, we will learn how to generalize this method to the case of a ``dilute'' 
system of composite bosons, for which the effective two-boson interaction depends on
frequency as well as on wave vector. 

We will eventually extrapolate the approximation thus obtained from the bosonic 
to the (fermionic) weak-coupling limit through the crossover region. 
In the latter region (where the parameter $k_{F} a_{F}$ exceeds unity), however, the 
``low-density''approximation has to be regarded merely as a scheme which interpolates
between the two controlled limits.
\subsection{Low-density approximation for true (point-like) bosons}

The standard argument to select the diagrams giving the leading contribution to 
the self-energy for a ``low-density'' Bose gas proceeds as 
follows.\cite{Beliaev,Popov-1,Popov-2}
Let $u({\mathbf q}_{1},{\mathbf q}_{2},{\mathbf q}_{3},{\mathbf q}_{4})$ be the 
(symmetrized) bosonic interaction potential, assumed to be vanishing for 
$|{\mathbf q}_{i}| \gapprox r_{o}^{-1}$ ($i=1,\cdots,4$), where $r_{o}$ is the 
range of the potential.
We shall also consider temperatures not too higher than the BE critical temperature,
so that we shall assume $T \sim \rho^{2/3}$.
Under these conditions, it turns out that \emph{every cycle in a diagram
contributes a factor\/} $T^{3/2} \sim \rho$.\cite{footnote-loop}

\begin{figure}
\narrowtext
\epsfysize=4.0in\hspace{0.2in} 
\epsfbox{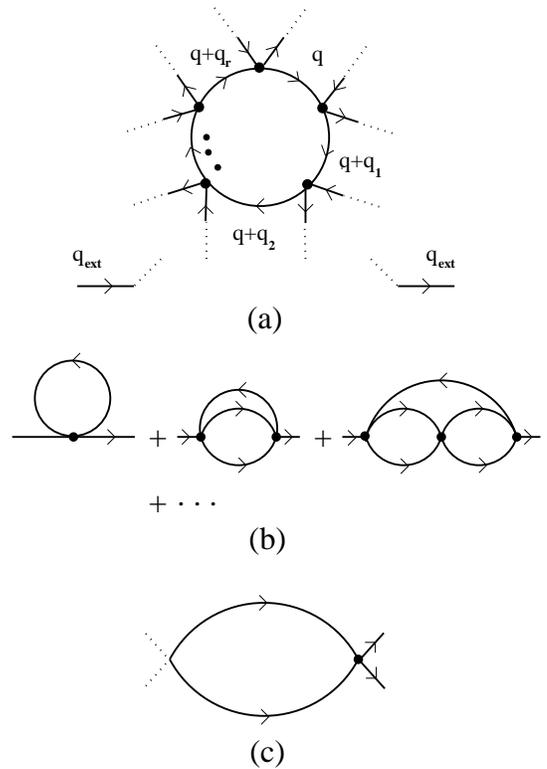}
\vspace{.2truecm}
\caption{\emph{(a)} Generic self-energy diagram 
for a system of \emph{true\/} bosons, containing one cycle (full lines and 
full circles here stand for the ``bare'' bosonic propagators and the (symmetrized) 
interaction potential, respectively, while dotted lines represent connections 
to the rest of the diagram); \emph{(b)} T-matrix approximation to the self-energy of 
true bosons; \emph{(c)} Elementary rung for the bosonic T-matrix.}         
\label{fig:bosloop}
\end{figure}

To show this, we consider the generic self-energy diagram of 
Fig.~\ref{fig:bosloop}a, containing one closed path constructed with ``bare''
bosonic propagators.
We perform explicitly the sum over the four-momentum $q$ running along the cycle.
Let $\xi_{B}({\mathbf q}) = {\mathbf q}^{2}/(2 m_{B}) - \mu_{B}$, where $m_{B}$ is
the bosonic mass and $\mu_{B}$ the bosonic chemical potential, respectively.
To the diagram of Fig.~\ref{fig:bosloop}a we then associate the expression

\begin{eqnarray}
& &{\mathcal L}({\mathbf q}_{ext}) = \sum_{P,Q}{\mathcal F}(P,Q)\sum_{q} 
U({\mathbf q};{\mathbf Q})\frac{1}{i\Omega_{\nu} - \xi_{B}({\mathbf q})}          \label{true-boson-loop}  \\ 
& \times & 
\frac{1}{i(\Omega_{\nu}+\Omega_{\nu_{1}}) - \xi_{B}({\mathbf q}+{\mathbf q}_{1})}
\cdots
\frac{1}{i(\Omega_{\nu}+\Omega_{\nu_{r}}) - \xi_{B}({\mathbf q}+{\mathbf q}_{r})}
                                                                 \nonumber
\end{eqnarray}

\noindent
where $U({\mathbf q};{\mathbf Q})$ stands for the product of all interaction potentials 
attached to the cycle (with all wave vectors other than ${\mathbf q}$ indicated
collectively by ${\mathbf Q}$) and ${\mathcal F}(P,Q)$ for the expression of the rest 
of the diagram (appropriate factors $(\beta {\mathcal V})^{-1}$ are understood to 
be associated with the four-momentum sums). 

The sum over the Matsubara frequency $\Omega_{\nu}$ can be readily performed in
Eq.~(\ref{true-boson-loop}), since the interaction potential for point-like bosons
is frequency independent. For this sum we obtain the result:

\begin{eqnarray}
& - & \frac{1}{e^{\beta \xi_{B}({\mathbf q})} - 1}
\frac{1}{\xi_{B}({\mathbf q}) + i\Omega_{\nu_{1}} 
                       - \xi_{B}({\mathbf q}+{\mathbf q}_{1})}\cdots
\nonumber\\
&\times&     \frac{1}{\xi_{B}({\mathbf q}) + i\Omega_{\nu_{r}}  
                       - \xi_{B}({\mathbf q}+{\mathbf q}_{r})}     
\nonumber \\
& - & 
\frac{1}{e^{\beta \xi_{B}({\mathbf q}+{\mathbf q}_{1})} 
-1}\frac{1}{\xi_{B}({\mathbf q}+{\mathbf q}_{1}) - i\Omega_{\nu_{1}} 
                       - \xi_{B}({\mathbf q})}\cdots\nonumber\\
&\times& \frac{1}{\xi_{B}({\mathbf q}+{\mathbf q}_{1}) + i(\Omega_{\nu_{r}}-\Omega_{\nu_{1}}) - \xi_{B}({\mathbf q}+{\mathbf q}_{r})}  
-\cdots   \,\, . \label{sum-over-loop}
\end{eqnarray}

\noindent
When integrating further over the wave vector ${\mathbf q}$ in 
Eq.~(\ref{true-boson-loop}) (and provided $\mu_{B}$ is negative), 
each Bose factor in Eq.~(\ref{sum-over-loop}) introduces a \emph{cutoff\/} 
proportional to $T^{1/2} \sim \rho^{1/3}$ about ${\mathbf q} + {\mathbf q}_{j}
\approx 0$ (with ${\mathbf q}_{j} = 0,{\mathbf q}_{1},\cdots,{\mathbf q}_{r}$).
This cutoff is much smaller than the range $r_{o}^{-1}$ of the bosonic interaction 
potential, owing to the ``low-density'' condition $\rho^{1/3} r_{o} \ll 1$.
We can then approximate ${\mathbf q} \approx 0$ everywhere in the integrand
associated with the first term of Eq.~(\ref{sum-over-loop}), ${\mathbf q} + 
{\mathbf q}_{1} \approx 0$ in the integrand associated with the second term of
Eq.~(\ref{sum-over-loop}), and so on, \emph{except\/} for the Bose factors that
retain the wave vector dependence indicated in Eq.~(\ref{sum-over-loop}).
We thus obtain the following approximate result for the sum over $q$ in the 
expression (\ref{true-boson-loop}):

\begin{eqnarray}
& - &   \frac{1}{i\Omega_{\nu_{1}} - \frac{{\mathbf q}_{1}^{2}}{2m_{B}}}   
\cdots
\frac{U(0;{\mathbf Q})}{i\Omega_{\nu_{r}} - 
\frac{{\mathbf q}_{r}^{2}}{2m_{B}}} 
\int\!\frac{d{\mathbf q}}{(2 \pi)^{3}}\frac{1}{e^{\beta \xi_{B}({\mathbf q})} -1}
\nonumber \\
& - &   \frac{1}{- i\Omega_{\nu_{1}} - \frac{{\mathbf q}_{1}^{2}}{2m_{B}}}  \cdots
        \frac{U(-{\mathbf q}_{1};{\mathbf Q})}
{i(\Omega_{\nu_{r}}-\Omega_{\nu_{1}}) - \frac{({\mathbf q}_{r}-
{\mathbf q}_{1})^{2}}{2m_{B}}}\nonumber\\    
&\times & \int\!\frac{d{\mathbf q}}{(2 \pi)^{3}}
        \frac{1}{e^{\beta \xi_{B}({\mathbf q}+{\mathbf q}_{1})} -1} 
-\cdots \label{integral-over-loop} 
\end{eqnarray}

\noindent 
where each integral equals the bosonic density $\rho_{B}=\rho/2$.
For generic values of the wave vectors ${\mathbf q}_{1},\cdots,{\mathbf q}_{r}$
and of the frequencies $\Omega_{\nu_{1}},\cdots,\Omega_{\nu_{r}}$ entering 
Eq.~(\ref{true-boson-loop}), each term in Eq.~(\ref{integral-over-loop}) is thus 
proportional to the density.
The presence of one cycle in Eq.~(\ref{true-boson-loop}) makes, in turn, 
the whole expression proportional to $\rho$.

It then follows that, for a ``low-density'' Bose system, the leading self-energy
diagrams contain the \emph{minimum number\/} of cycles, like the diagrams shown
in Fig.~\ref{fig:bosloop}b, which constitutes the so-called bosonic T-matrix
approximation.\cite{Beliaev,Popov-1,Popov-2}
It is interesting to contrast the behaviour of the cycle (whereby the
Bose factor introduces a cutoff proportional to $T^{1/2} \sim \rho^{1/3}$) with
the behaviour of an elementary rung (drawn in Fig.~\ref{fig:bosloop}c) in the 
diagrammatic representation of the bosonic T-matrix, with one potential vertex and 
two bosonic propagators running in the \emph{same\/} direction. 
In this case, the cutoff originates from the potential and \emph{not\/} from the 
Bose factor.
Let us, in fact, assume, for simplicity, that the bosonic interaction potential is
(approximately) constant ($\sim u(0)$) up to the cutoff $r_{o}^{-1}$. 
It can then be readily shown that, at low enough temperature, the dominant contribution 
to the diagram of Fig.~\ref{fig:bosloop}c is given by $u(0) \, r_{o}^{-1}$, in
analogy with the result (\ref{I-pp}) for fermions.
No density factor thus arises in this case.

To generalize the above results to a ``dilute'' system of \emph{composite\/}
bosons, it is essential to take into account the \emph{frequency dependence\/} of
the effective two-boson interaction (\ref{two-body-potential}), which makes the
calculation of the analogue of the expression (\ref{true-boson-loop}) somewhat 
more involved. We will, however, show in the next subsection that this quantity 
remains proportional to $\rho$.
\subsection{Low-density approximation for composite bosons}

\begin{figure}
\narrowtext
\epsfysize=4.0in 
\epsfbox{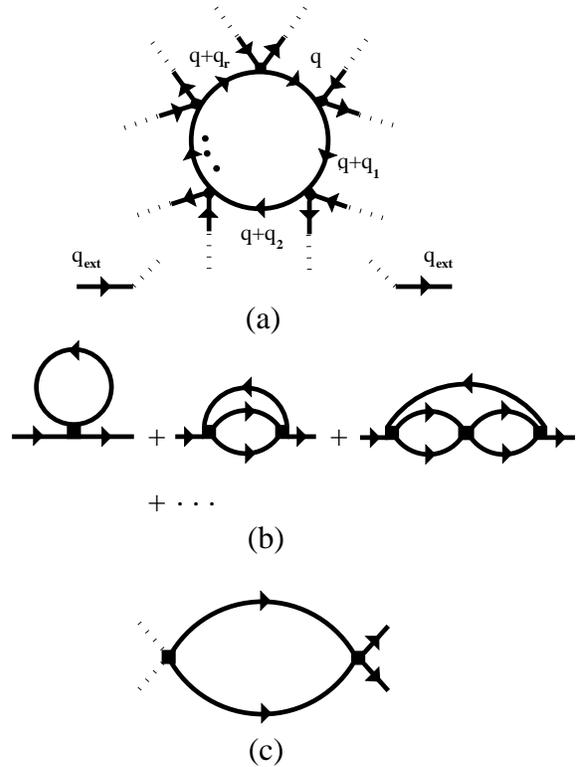}
\vspace{.2truecm}
\caption{ \emph{(a)} Generic self-energy 
diagram for a system of \emph{composite\/} bosons, containing one cycle 
(thick lines and full squares stand for the ``bare'' propagators for composite bosons and 
the effective two-boson interaction, respectively, according 
to the conventions 
of Fig.~\ref{fig:cobos}, while dotted lines represent connections 
to the rest of the diagram); \emph{(b)} T-matrix approximation to the self-energy 
of composite bosons; \emph{(c)} Elementary rung for the T-matrix of composite bosons.}         
\label{fig:comboslo}
\end{figure}

The generic self-energy diagram containing one cycle and associated with the propagator 
$<b^*(q) b(q)>_{S_{\mathrm{eff}}}$ for composite bosons is shown in 
Fig.~\ref{fig:comboslo}a. 
Here, thick lines and squares represent the ``bare'' propagator 
$<b^*(q) b(q)>_{S^{(2)}_{\mathrm{eff}}}$ of Eq.~(\ref{bare bb}) for composite bosons 
and the effective two-boson interaction given by Eq.~(\ref{two-body-potential}), 
respectively, according to the conventions of Fig.~\ref{fig:cobos}.
[Recall from subsection IIB that only the two-boson interaction needs to be taken 
into account in the strong-coupling limit.]

The expression of the ``bare'' propagator $<b^*(q) b(q)>_{S^{(2)}_{\mathrm{eff}}}$ is 
obtained by inserting in Eqs.~(\ref{bare bb}) and (\ref{p-p ladder}) the expression 
(\ref{R-pp-sc}) for $R_{pp}$, that holds specifically in the strong-coupling limit 
(whereby $\beta \mu \rightarrow - \infty$).
Writing $\sqrt{E_{\alpha}} \mathrm{sgn}(\mathrm{Im}E_{\alpha}) = i \sqrt{- E_{\alpha}}$ 
and recalling the definition (\ref{E-alpha}) for $E_{\alpha}$, 
we obtain:

\begin{eqnarray}
<b^*(q) b(q)>_{S^{(2)}_{\mathrm{eff}}} & = & \frac{\beta}{{\mathcal V}} \frac{-1}
{\frac{m}{4 \pi a_{F}} \, - \, \frac{m^{3/2}}{4 \pi} 
\sqrt{-i\Omega_{\nu} + \frac{{\mathbf q}^{2}}{4 m} - 2\mu}}      \nonumber \\
& = & \frac{D(q)}
{i\Omega_{\nu} - \left(\frac{{\mathbf q}^{2}}{4m} - \mu_{B}\right)}  \,\,  ,
                                                                \label{general-pp-sc}
\end{eqnarray}

\noindent
where we have used the definition $\mu_{B} = 2\mu + \epsilon_{o}$ for the bosonic
chemical potential and introduced the notation \cite{footnote-residue}

\begin{equation}
D(q) \, = \, - \, \frac{\beta}{{\mathcal V}} \, \frac{4 \pi}{m^{2} a_{F}} \left(1 \, + \,
\sqrt{1 \, + \, \frac{-i\Omega_{\nu} +
\frac{{\mathbf q}^{2}}{4 m} - \mu_{B}}{\epsilon_{o}}} \, \right)    \, .     \label{residue}
\end{equation}

\noindent
Replacing further $i\Omega_{\nu}$ by the complex frequency $z$, we note that, near 
the position $z = {\mathbf q}^{2}/(4m) - \mu_{B}$ of the pole in 
Eq.~(\ref{general-pp-sc}), $D(q)$ of Eq.~(\ref{residue}) reduces to 
($- \beta / {\mathcal V}$ times) the constant value $8 \pi/(m^{2} a_{F})$ entering 
Eq.~(\ref{pp-sc}) and thus plays the role of a residue.
Away from this pole, $D(q)$ has a cut along the positive real axis for 
Re($z$) $\ge - 2\mu$.

To retain the structure of Eq.~(\ref{true-boson-loop}) as close as possible
(with $m_{B}=2m$ in the ``bare'' bosonic propagators), it is convenient to transfer the 
factors $D(q)$ from the ``bare'' propagators $<b^*(q) b(q)>_{S^{(2)}_{\mathrm{eff}}}$
to the effective two-boson interactions, which these propagators are joined to.
Specifically, every propagator $<b^*(q) b(q)>_{S^{(2)}_{\mathrm{eff}}}$ associated with
an ``internal'' line of a diagram tranfers a factor $\sqrt{D(q)}$ to each of the
two interaction vertices it is joined to, while each propagator associated with one of 
the two ``external'' lines transfers one factor $\sqrt{D(q)}$ to the single interaction 
it is joined to and assigns the remaining factor $\sqrt{D(q)}$ as a proportionality 
factor to the ``full'' propagator, which the diagram is meant to represent.
In this way, the ``full'' propagator acquires the same overall factor $D(q)$
of the ``bare'' propagator (\ref{general-pp-sc}) and the effective two-boson
interaction of Eq.~(\ref{two-body-potential}) is multiplied by
$\sqrt{D(q_{1})} \sqrt{D(q_{2})} \sqrt{D(q_{3})} \sqrt{D(q_{4})}$.
We are then led to \emph{rescaling\/} the effective two-boson interaction as follows 

\begin{eqnarray}
& &2 u_{2}(q_{1} \dots q_{4}) \label{rescaled-two-body-potential}  \\ 
& &\equiv \tilde{u}_{2}(q_{1} \dots q_{4}) 
\sqrt{D(q_{1})} \sqrt{D(q_{2})} \sqrt{D(q_{3})} \sqrt{D(q_{4})}
\nonumber  
\end{eqnarray}

\noindent
with $\tilde{u}_{2}(q_{1} \dots q_{4})$ given by Eq.~(\ref{two-body-potential}).
[Note that the above definition accounts, in particular, for the difference between
$\tilde{u}_{2}(0)$ and $u_{2}(0)$ in Eq.~(\ref{u2-tilde-limit}).]

To the generic self-energy diagram of Fig.~\ref{fig:comboslo}a,
containing one cycle of ``bare'' propagators for composite bosons, 
we thus associate an expression similar to (\ref{true-boson-loop}), namely,

\begin{eqnarray}
& &{\mathcal L}({\mathbf q}_{ext}) = \sum_{P,Q} {\mathcal F}(P,Q) 
\sum_{q}U(q;Q)
\frac{1}{i\Omega_{\nu} - \xi_{B}({\mathbf q})}\label{composite-boson-loop}\\
& &\times\frac{1}{i(\Omega_{\nu}+\Omega_{\nu_{1}}) - \xi_{B}({\mathbf q}+{\mathbf q}_{1})}
\cdots\frac{1}{i(\Omega_{\nu}+\Omega_{\nu_{r}}) - \xi_{B}({\mathbf q}+{\mathbf q}_{r})}\nonumber
\end{eqnarray}

\noindent
where $U(q;Q)$ now stands for the product of all rescaled effective two-boson 
interactions $2 u_{2}$ attached to the cycle.
Although the sum over the Matsubara frequency $\Omega_{\nu}$ in 
Eq.~(\ref{composite-boson-loop}) cannot be performed explicitly as we have done in
Eq.~(\ref{true-boson-loop}), it can be conveniently transformed into a contour
integration as in Eq.~(\ref{Sigma-T-matrix-B-0}). 
We thus obtain for this frequency sum:

\begin{eqnarray}
& &\oint_{C} \! \frac{dz}{2 \pi i} \frac{1}{z - \xi_{B}({\mathbf q})}
\frac{1}{z+i\Omega_{\nu_{1}} - \xi_{B}({\mathbf q}+{\mathbf q}_{1})}
\cdots\nonumber\\
& &\times\frac{1}{z+i\Omega_{\nu_{r}} - \xi_{B}({\mathbf q}+{\mathbf q}_{r})} \,
\frac{U({\mathbf q},z;Q)}{e^{\beta z} - 1} .\label{integral-z}
\end{eqnarray}

To proceed further, we need to spell out the analytic properties of 
$U({\mathbf q},z;Q)$.
From Fig.~\ref{fig:comboslo}a one realizes that \emph{each\/} factor
$\tilde{u}_{2}$ entering the expression for $U$ in Eq.~(\ref{composite-boson-loop}) 
has the following structure [cf. Eq.~(\ref{two-body-potential})]:

\begin{eqnarray}   
& &\tilde{u}_{2}(q+q_{j},q_{a},q_{b},q+q_{j-1})  
= \delta_{q_{j}+q_{a},\,\, q_{j-1}+q_{b}} 
     \, \left(\frac{{\mathcal V}}{\beta}\right)^{2} \, 
     \frac{2}{\beta {\mathcal V}}\nonumber\\ 
& &\times \sum_k
\frac{1}{\epsilon(-k)\epsilon(k+q_{a})\epsilon(-k+q_{j}-q_{j-1})\epsilon(k+q+q_{j-1})}
\label{two-body-potential-ad-hoc}
\end{eqnarray}

\noindent
where $q_{a}$ and $q_{b}$ belong to the set $Q$ of Eq.~(\ref{composite-boson-loop}).
In other words, the cycle variable $q$ in Eq.~(\ref{two-body-potential-ad-hoc}) appears
\emph{only\/} in the last factor $\epsilon(k+q+q_{j-1})^{-1}$.  
Viewed as a function of the complex variable $z$ of Eq.~(\ref{integral-z}),
each factor $\tilde{u}_{2}$ has thus a \emph{cut\/} along the positive real axis for 
Re($z$) $\ge - \mu$.
This property, in turn, remains true for the rescaled effective two-boson potential
$u_{2}$ of Eq.~(\ref{rescaled-two-body-potential}), since we have already determined 
that the factor $D(q)$ has itself a cut along the positive real axis for 
Re($z$) $\ge - 2\mu$.

It is thus clear that the contribution of the singularities of $U({\mathbf q},z;Q)$ to
the integral of Eq.~(\ref{integral-z}) is strongly suppressed by the presence of the 
Bose factor $(\exp\{\beta z\} - 1)^{-1}$ therein, since $\beta |\mu| \gg 1$ in the
strong-coupling limit.
In this limit, the contribution of these singularities can then be neglected, 
with the result that the value of the expression (\ref{composite-boson-loop}) is 
proportional to the bosonic density $\rho_{B}$, by the very argument discussed in the 
previuos subsection for point-like bosons.

We conclude that, for a system of composite bosons in the ``low-density'' limit, the 
leading diagrams contain just \emph{one cycle\/}, like the ones shown in
Fig.~\ref{fig:comboslo}b.
In analogy with the corresponding diagrams of Fig.~\ref{fig:bosloop}b for 
point-like bosons, we shall refer to these diagrams as the T-matrix approximation 
for composite bosons.

That in the strong-coupling limit all diagrams of Fig.~\ref{fig:comboslo}b are of the 
\emph{same order\/} (in the original fermionic parameter $k_{F} a_{F}$ - 
cf.\ Ref.\ 14) can be further verified as follows.
Consider the elementary rung (drawn in Fig.~\ref{fig:comboslo}c)
in the diagrammatic representation of the T-matrix for composite bosons.
We know that the interaction vertex $u_{2}$ scales as $|\mu|^{-3/2} \times a_{F}^{-2}$
in the strong-coupling limit, where the factor $|\mu|^{-3/2}$ stems 
from the behaviour of $\tilde{u}_{2}$ (cf.\ the comment after 
Eq.~(\ref{three-body-potential})) and the factor $a_{F}^{-2}$ from the presence of 
two powers of $D$ in the definition  (\ref{rescaled-two-body-potential}) of the 
rescaled interaction vertex $u_{2}$ (cf.\ also Eq.~(\ref{residue})).
By the argument made for point-like bosons at the end of the previous subsection, 
the cutoff on the four-momentum of the rung is provided by the interaction 
vertex since the two ``bare'' propagators for composite bosons run in the same
direction.
This four-momentum sum is then proportional to the cutoff of the interaction
vertex, which we know from Ref.\ 16 to be given by $a_{F}^{-1}$.
Putting all factors together and recalling that $|\mu|^{-1/2} \propto a_{F}$
in the strong-coupling limit, we obtain that the rung of 
Fig.~\ref{fig:comboslo}c scales like $|\mu|^{-3/2} \times a_{F}^{-2}
\times a_{F}^{-1} \sim constant$.
This result implies that the order of the diagram in the parameter $k_{F}a_{F}$ is not
altered by the insertion of an \emph{arbitrary number\/} of these rungs.

The diagrams of Fig.~\ref{fig:comboslo} have been drawn in terms of propagators and 
interaction vertices for composite bosons.
When extrapolating these diagrams from the bosonic to the fermionic (weak-coupling) 
limit through the crossover region, however, it will be useful to draw them
also in the alternative representation in terms of the constituent fermions,
via the correspondence rules of Fig.~\ref{fig:cobos}.
In Appendix B we will examine in detail which fermionic diagrams are associated 
with the T-matrix approximation for composite bosons.
In this way, we shall relate the value of the symmetry factors of the bosonic 
diagrammatic structure with the number of independent diagrams in the corresponding 
fermionic diagrammatic structure.
We shall also show that in the weak-coupling limit, where the composite-boson
propagator becomes a constant to leading-order in $k_{F} a_{F}$ (cf.
Eq.~(\ref{pp-wc})), our generalized T-matrix approximation reduces correctly
to the Galitskii approximation for the self-energy of a ``low-density'' Fermi gas.
 
\subsection{Coupled equations defining the generalized T-matrix approximation 
for the BE-BCS crossover}

Once the self-energy diagrams for composite bosons in the ``low-density'' limit
have been selected according to the above prescriptions, there remains to determine 
the analytic expression of these diagrams.
To this end, it is convenient to stick with the bosonic representation and
write down the expression for the bosonic propagator $<b^*(q) b(q)>_{S_{\mathrm{eff}}}$
with the self-energy insertions of Fig.~\ref{fig:comboslo}b, making 
use of the standard bosonic diagrammatic rules.\cite{Popov-1,Popov-2}
Recalling the general relation (\ref{A-final}) of Appendix A for the ``full'' bosonic
propagator (as well as its counterpart (\ref{bare bb}) for the ``bare'' 
bosonic propagator), we obtain:

\begin{eqnarray}
& &\frac{\beta}{{\mathcal V}} \Gamma(q) = \frac{\beta}{{\mathcal V}} \left\{
\Gamma_{o}(q)  - \Gamma_{o}(q)^{2} \frac{1}{\beta {\mathcal V}} \sum_{q'} 
\Gamma_{o}(q')\right.\nonumber\\
& &\times \left[{\mathcal S}_{1}2\bar{u}_{2}(q',q,q,q')
-\frac{{\mathcal S}_{2}}{\beta{\mathcal V}} \sum_{q_{1}} 
2 \bar{u}_{2}(q',q,q_{1},q+q'-q_{1})   
\right. \nonumber\\
& &\times\Gamma_{o}(q_{1})\Gamma_{o}(q+q'-q_{1})2\bar{u}_{2}(q+q'-q_{1},q_{1},q,q') 
\nonumber  \\
& &+\frac{{\mathcal S}_{3}}{(\beta {\mathcal V})^2}\sum_{q_{1},q_{2}} 
2 \bar{u}_{2}(q',q,q_{2},q+q'-q_{2}) \Gamma_{o}(q_{2})\Gamma_{o}(q+q'-q_{2})
\nonumber  \\
& &\times 2 \bar{u}_{2}(q+q'-q_{2},q_{2},q_{1},q+q'-q_{1})  
\Gamma_{o}(q_{1}) \, \Gamma_{o}(q+q'-q_{1})\nonumber\\
& &\times 2 \bar{u}_{2}(q+q'-q_{1},q_{1},q,q')
- \left. \left. \cdots \phantom{\frac{1}{1}} \right] + \cdots \right\}   
\; . \label{bosonic-Green-function}
\end{eqnarray}

\noindent
In this expression, ${\mathcal S}_{L} = 2^{1-L} (L=1,2,\cdots)$ is the 
\emph{symmetry factor\/} associated with the bosonic self-energy diagram of 
Fig.~\ref{fig:comboslo}b containing $L$ bosonic interaction vertices
[cf. Appendix B], $\bar{u}_{2}$ is proportional to the effective two-boson
interaction of Eq.~(\ref{two-body-potential})

\begin{eqnarray}
& &\bar{u}_{2}(q_{1} \dots q_{4}) \, = \,  \frac{1}{\beta {\mathcal V}} \, 
\sum_{k} \nonumber\\
& &\times\frac{1}{\epsilon(-k)\epsilon(k+q_2)\epsilon(-k+q_1-q_4)
\epsilon(k+q_4)} \; ,
\label{two-body-potential-bar}     
\end{eqnarray}

\noindent
and $\Gamma_{o}$ is the ``bare'' particle-particle ladder of Fig.~\ref{fig:pplad}a, 
whose complete expression

\begin{eqnarray}
& &\Gamma_{o}(q) = - \left\{ \frac{m}{4 \pi a_{F}} + \right.
\int \! \frac{d{\mathbf k}}{(2 \pi)^{3}}\nonumber\\ 
& &\times\left. \left[\frac{\tanh(\beta \xi({\mathbf k})/2) 
+\tanh(\beta \xi({\mathbf k-q})/2)}{2(\xi({\mathbf k})+\xi({\mathbf k-q})-i
\Omega_{\nu})} 
- \frac{m}{{\mathbf k}^{2}} \right] \right\}^{-1}
\label{most-general-pp-sc}
\end{eqnarray}

\noindent
holds irrespective of the value of $a_{F}$ and reduces in particular to the form
(\ref{general-pp-sc}) in the strong-coupling limit $\beta \mu \rightarrow  - \infty$.

Note that the factor $2$, which entered the definition (\ref{two-body-potential}) 
for $\tilde{u}_{2}$ but has been removed from the definition 
(\ref{two-body-potential-bar}) for $\bar{u}_{2}$, 
multiplies now each $\bar{u}_{2}$ in Eq.~(\ref{bosonic-Green-function}) explicitly.
These factors of $2$ combine with the symmetry factors ${\mathcal S}_{L}$ 
in Eq.~(\ref{bosonic-Green-function}), resulting in an overall factor of $2$
multiplying the whole expression within brackets therein.
This expression corresponds then to the following integral equation

\begin{eqnarray}
& &\bar{t}_{B}(q_{1},q_{2},q_{3},q_{4}) =\bar{u}_{2}(q_{1},q_{2},q_{3},q_{4}) 
\nonumber\\ 
& &-\frac{1}{\beta {\mathcal V}} \, \sum_{q_{5}} \,
 \bar{u}_{2}(q_{1},q_{2},q_{5},q_{1}+q_{2}-q_{5}) 
\Gamma_{o}(q_{5}) \, \Gamma_{o}(q_{1}+q_{2}-q_{5})\nonumber\\
& &\times \bar{t}_{B}(q_{1}+q_{2}-q_{5},q_{5},q_{3},q_{4}) 
\label{bosonic-t-matrix}
\end{eqnarray}

\noindent
with $q_{1}+q_{2}=q_{3}+q_{4}$, provided we set $q_{1}=q_{4}=q'$ and $q_{2}=q_{3}=q$.
Equation (\ref{bosonic-t-matrix}) identifies the \emph{generalized\/} T-\emph{matrix\/} 
\emph{for composite bosons\/}, in analogy  with the ordinary T-matrix for true 
bosons.\cite{Beliaev,Popov-1,Popov-2}
It is thus clear from Eq.~(\ref{bosonic-Green-function}) that the quantity

\begin{equation}
\Sigma_{B}^{(t)}(q) \, = \, - \, \frac{2}{\beta {\mathcal V}} \, \sum_{q'} \,
\Gamma_{o}(q') \, \bar{t}_{B}(q',q,q,q')                    \label{bosonic-self-energy}
\end{equation}

\noindent
represents the T-matrix approximation to the self-energy for composite bosons, in
terms of which Eq.~(\ref{bosonic-Green-function}) acquires the characteristic form
of a Dyson's equation:

\begin{eqnarray}
\Gamma(q) & = & \Gamma_{o}(q) \, + \, 
\Gamma_{o}(q) \, \Sigma_{B}^{(t)}(q) \, \Gamma_{o}(q) \, + \, \cdots  \nonumber    \\
& = & \frac{\Gamma_{o}(q)}{1 \, - \, \Sigma_{B}^{(t)}(q) \, \Gamma_{o}(q)} \, = \, 
\frac{1}{\frac{1}{\Gamma_{o}(q)} \, - \, \Sigma_{B}^{(t)}(q)}  \,\, .
                                                            \label{bosonic-GF-final}
\end{eqnarray}

With the above expression for the particle-particle ladder (which, by construction,
correctly describes a system of ``low-density'' composite bosons in the strong-coupling
limit), we can obtain the fermionic self-energy of interest in analogy to 
Eq.~(\ref{Sigma-T-matrix}), by joining the incoming and outgoing arrows of the 
particle-particle ladder $\Gamma$ with a single-particle fermionic Green's function 
in the only possible way. We then write:

\begin{equation}
\Sigma_{F}(k) \, = \, - \, \frac{1}{\beta {\mathcal V}} \, \sum_{k'} \,
\Gamma(k+k') \,\, {\mathcal G}^{o}(k')                  \label{Sigma-T-matrix-final}
\end{equation}

\noindent
with $\Gamma$ given by Eq.~(\ref{bosonic-GF-final}) and where ${\mathcal G}^{o}$
is the ``bare'' single-particle fermionic Green's function.
The self-energy (\ref{Sigma-T-matrix-final}) has in turn to be inserted into the 
fermionic Dyson's equation, to yield the full single-particle fermionic Green's function 
${\mathcal G}$.
Eventual extrapolation from the strong- to the weak-coupling limit through the 
crossover region requires us to eliminate the chemical potential in favor of the
particle density $\rho$, by evaluating

\begin{equation}
\rho \, = \, \frac{2}{\beta {\mathcal V}} \, \sum_{k} \, e^{i\omega_{n}\eta} \,
               {\mathcal G}(k)                                       \label{density}
\end{equation}

\noindent
where $\eta$ is a positive infinitesimal.

Besides the explicit ${\mathcal G}^{o}$ in Eq.~(\ref{Sigma-T-matrix-final}), also 
\emph{all\/} single-particle fermionic Green's functions entering the expression 
(\ref{bosonic-GF-final}) for $\Gamma$ are meant to be ``bare'' ones, in analogy 
to the original approach by Galitskii.\cite{footnote-6}
We expect, in fact, that the inclusion of self-consistency in the explicit 
single-particle fermionic Green's function of Eq.~(\ref{Sigma-T-matrix-final}) 
should not be essential to represent correctly the fermionic self-energy,
either in the strong-coupling limit (where, on physical grounds, it is rather the
bosonic propagator that needs to be represented correctly) or in the weak-coupling
limit (where self-consistency drops out anyway for a ``low-density'' Fermi system).
[In addition, in the crossover region (where our ``low-density''approximation 
merely interpolates between the two controlled limits) inclusion of self-consistency
should not be of primary concern.] 
By the same token, inclusion of self-consistency in the single-particle fermionic
Green's functions entering the expression (\ref{bosonic-GF-final}) for $\Gamma$
would yield contributions of higher order in the small parameter $k_{F}a_{F}$ with 
respect to the ones retained (both in the weak- and strong-coupling limits).

The question of self-consistency in the single-particle fermionic Green's function
is also connected with the general requirement for an approximation to the fermionic
self-energy to be ``conserving'' in the Baym-Kadanoff sense.\cite{BK,Baym}
According to this requirement, once an approximation has been selected by physical
considerations \emph{at the level of\/} the fermionic two-particle Green's function 
(in the particle-particle channel), one should in principle construct a functional
$\Phi$ such that the fermionic self-energy itself is, in turn, obtained by taking the
functional derivative of $\Phi$ with respect to the self-consistent ${\mathcal G}$.
It is obvious that this procedure yields several \emph{additional\/} self-energy terms,
over and above the one obtained by joining the incoming and outgoing arrows of the 
particle-particle ladder with a single-particle fermionic Green's function, as we 
did in Eq.~(\ref{Sigma-T-matrix-final}). 
It is, however, also clear that, \emph{by construction\/}, these additional terms
would be of higher-order in the ``low-density'' parameter, and can thus be consistently
neglected.

A complete numerical evaluation of Eqs.~(\ref{two-body-potential-bar})-(\ref{density}) 
exceeds the purposes of the present paper. 
In the next Section we shall calculate the scattering length for composite
bosons in the strong-coupling limit, as a degenerate case of the T-matrix
given by Eq.~(\ref{bosonic-t-matrix}).
This calculation will ensure us that the diagrams of Fig.~\ref{fig:comboslo}b, 
beyond first order in the interaction potential for composite bosons, give 
contributions of the \emph{same order\/} of magnitude as the first-order 
(Hartee-Fock) diagram.
In addition, it will turn out that the series of diagrams depicted in 
Fig.~\ref{fig:comboslo}b does \emph{not\/} converge, making any truncation of the
series not appropriate.
For this reason, it is essential to solve the complete integral equation associated
with this series of diagrams to get a correct description of the 
strong-coupling limit.
\section{Numerical results for the scattering length of composite bosons}

The \emph{scattering length\/} $a$ characterizes the low-energy collisions for 
the scattering from an ordinary potential.\cite{Fermi}
For the mutual scattering of two particles (each of mass $M$), $a$ can be expressed 
by the relation $t(0) = 4\pi a/M$ in terms of the ordinary T-matrix $t(0)$ in the 
limit of vanishing wave vector.
In particular, within the Born approximation $t(0)$ is replaced by the
Fourier trasform $u(0)$ of the inter-particle potential for vanishing wave vector.

In a similar way, we shall \emph{define\/} the scattering length $a_{B}$ for composite 
bosons (each of mass $2m$) in the strong-coupling limit and for vanishing density, 
by setting

\begin{equation}
t_{B}(0) \, = \, \frac{4 \pi}{2m} \, a_{B}                       \label{tB-vs-aB}
\end{equation}

\noindent
where $t_{B}(0) = (8 \pi/(m^{2} a_{F}))^{2} \, \bar{t}_{B}(0)$ and
$\bar{t}_{B}(0) \equiv \bar{t}_{B}(0,0,0,0)$ [cf. Eqs.~(\ref{u2-tilde-limit}),
(\ref{two-body-potential}), and (\ref{two-body-potential-bar})].
This quantity is expected to be important for the calculation of the bosonic
self-energy (\ref{bosonic-self-energy}), insofar as the generalized T-matrix for
composite bosons therein depends weakly on its arguments.

To lowest order in the effective interaction for composite bosons,
we can replace $t_{B}(0)$ by $u_{2}(0)$ and write

\begin{equation}  
u_{2}(0) \, = \, \frac{4 \pi}{2m} \, a^{\mathrm{Born}}_{B}        \label{u2-vs-aB-Born}
\end{equation}

\noindent
within the Born approximation.
Comparison with Eq.~(\ref{u2-limit}) yields then the value 
$a^{\mathrm{Born}}_{B} = 2 \, a_{F}$, as anticipated by Eq.~(\ref{a-Born}). 
Note also that the replacement of $t_{B}$ by $u_{2}$ (or, equivalently,
of $\bar{t}_{B}$ by $\bar{u}_{2}$) corresponds to the bosonic Hartree-Fock 
approximation, whereby \emph{only\/} the diagram at the left in 
Fig.~\ref{fig:comboslo}b is considered instead of the whole sequence.

To calculate $\bar{t}_{B}(0)$, it is convenient to determine first the auxiliary
quantity $\bar{t}_{B}(q,-q,0,0)$ by solving the following \emph{closed-form\/} equation 

\begin{eqnarray}
& &\bar{t}_{B}(q,-q,0,0) = \bar{u}_{2}(q,-q,0,0) \label{bosonic-t-matrix-0}\\ 
& & - \frac{1}{\beta {\mathcal V}} \, \sum_{q'} \,
     \bar{u}_{2}(q,-q,q',-q')\Gamma_{o}(q') \, \Gamma_{o}(-q') \,
     \bar{t}_{B}(q',-q',0,0)   \,\, , \nonumber 
\end{eqnarray}

\noindent
which is obtained from Eq.~(\ref{bosonic-t-matrix}) by setting $q_{1}=-q_{2}=q$ and
$q_{3}=q_{4}=0$.
This integral equation can be solved by standard numerical methods, e.g., by
reverting it to the solution of a system of coupled linear equations.

Before embarking into this numerical calculation, we can obtain a preliminary 
estimate of the value of $\bar{t}_{B}(0)$ with limited effort, by neglecting the 
four-vector dependence of $\bar{t}_{B}$ as well as the frequency dependence of 
$\bar{u}_{2}$ on the right-hand side of Eq.~(\ref{bosonic-t-matrix-0}).
We thus write approximately:

\begin{eqnarray}
& &\left( \frac{\bar{t}_{B}(0)}{\bar{u}_{2}(0)} \right)^{-1} =  
\left( \frac{t_{B}(0)}{u_{2}(0)} \right)^{-1}\nonumber\\ 
& &\simeq 
1 + \int \! \frac{d{\mathbf q}}{(2 \pi)^{3}}  
             \bar{u}_{2}({\mathbf q},-{\mathbf q},0,0) 
             \frac{1}{\beta} \sum_{\Omega_{\nu}}
             \Gamma_{o}(q) \Gamma_{o}(-q) \nonumber  \\
& & = 1 - \int \! \frac{d{\mathbf q}}{(2 \pi)^{3}}  
             u_{2}({\mathbf q},-{\mathbf q},0,0)
             \frac{1}{\beta} \sum_{\Omega_{\nu}} 
             \frac{1}{i\Omega_{\nu}-\frac{{\mathbf q}^{2}}{4m}}  
             \frac{1}{i\Omega_{\nu}+\frac{{\mathbf q}^{2}}{4m}}         
\nonumber \\   
& & = 1 + \int \! \frac{d{\mathbf q}}{(2 \pi)^{3}}  
             u_{2}({\mathbf q},-{\mathbf q},0,0) 
             \frac{1}{\frac{{\mathbf q}^{2}}{2m}}   \,\, ,\label{t00}
\end{eqnarray}

\noindent
where the last result holds to the leading order in the density.
Recalling further from Ref.\ 16 that

\begin{equation}
u_{2}({\mathbf q},-{\mathbf q},0,0) \, = \, u_{2}(0) \,\, 
      \frac{4}{4 + (|{\mathbf q}| a_{F})^{2}}                        \label{u-2-right}
\end{equation}

\noindent
in the strong-coupling limit, we obtain for the integral on the right-hand side 
of Eq.~(\ref{t00}) the following value (in three dimensions)

\begin{equation}
\int \! \frac{d{\mathbf q}}{(2 \pi)^{3}} \,\, u_{2}({\mathbf q},-{\mathbf q},0,0) \,\,
\frac{1}{\frac{{\mathbf q}^{2}}{2m}} \, = \, u_{2}(0) \, \frac{m}{\pi a_{F}}
\, = \, 4                                                      \label{value-integral}
\end{equation}

\noindent
where use has been made of the result (\ref{u2-limit}).
Comparison with Eqs.~(\ref{tB-vs-aB}) and (\ref{u2-vs-aB-Born}) yields eventually

\begin{equation}
\frac{a_{B}}{a_{B}^{\mathrm{Born}}} \, = \, \frac{t_{B}(0)}{u_{2}(0)} 
\, \simeq  \, \frac{1}{5}   \,\, ,                                  \label{1-5}
\end{equation}

\noindent
which implies that the contribution of the series diagrams depicted in  
Fig.~\ref{fig:comboslo}b is of the \emph{same order of magnitude\/}
as the contribution of the first-order (Hartree-Fock) diagram therein.

The above approximate calculation suggests us to consider in more detail the scattering
problem for two true bosons (each of mass $M$), mutually interacting via a potential 
of the form (\ref{u-2-right}), namely,

\begin{equation}
u({\mathbf q}) \, = \, \frac{8 \pi a_{F}}{M} \, \frac{\gamma}{1 + \tilde{q}^{2}/4}
                                                            \label{u-true-bosons}
\end{equation}

\noindent
where $\tilde{q} = |{\mathbf q}| a_{F}$ and  $\gamma$ is a dimensionless coupling 
constant ($\gamma = 1$ and $M=2m$ for the potential (\ref{u-2-right})).
Solving for this simplified scattering problem will, in fact, be instructive to obtain 
the solution of the original scattering problem (\ref{bosonic-t-matrix-0}) for composite
bosons, since it will 
\emph{(i)} suggest a nontrivial approximation to be carried over to the original 
equation (\ref{bosonic-t-matrix-0}), and 
\emph{(ii)} assess whether the region of interest ($\gamma \approx 1$) belongs to 
the perturbative or nonperturbative regime of the scattering integral equation.

To this end, we recall the equation satisfied by the T-matrix for two-body scattering.
In particular, it is sufficient to consider the following degenerate form

\begin{equation}
t({\mathbf q},0) \, = \,  u({\mathbf q}) \, - \, \int\!\frac{d{\mathbf q'}}{(2 \pi)^{3}}
\, \frac{u({\mathbf q} - {\mathbf q'}) \, t({\mathbf q'},0)}
        {\frac{{\mathbf q}^{'2}}{M}} \, \, ,
                                                           \label{bosonic-t-matrix-1}
\end{equation}

\noindent
which resembles Eq.~(\ref{bosonic-t-matrix-0}) for composite bosons but lacks its 
dependence on Matsubara frequencies.
As already noted, the scattering length is related to $t(0) \equiv t(0,0)$ by the
relation $a = t(0) M/(4 \pi)$.
For a spherically symmetric potential, $t({\mathbf q},0) = t(|{\mathbf q}|,0)$
and the angular integral in Eq.~(\ref{bosonic-t-matrix-1}) can be readily performed.
The remaining integral over $|{\mathbf q}|$ can be suitably discretized over
a mesh, until convergence is achieved for the desired value $t(0)$ (200 mesh points 
have proved sufficient to get a $1\%$ accuracy in the scattering length).

\begin{figure}
\narrowtext
\epsfxsize=3.3in 
\epsfbox{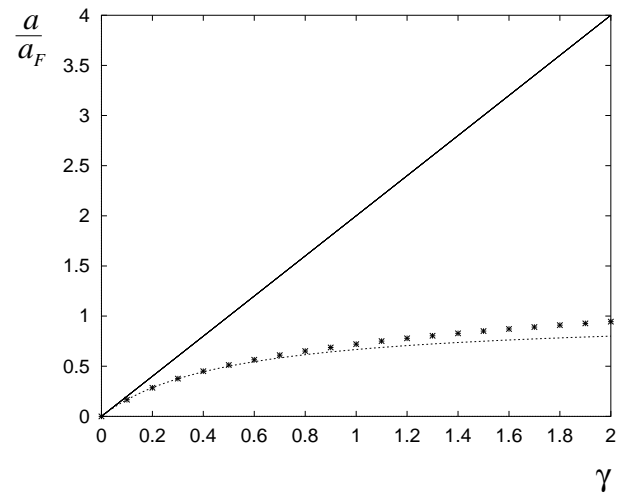}
\vspace{.2truecm}
\caption{Scattering length $a$ (in units of $a_{F}$)
vs. the coupling strength $\gamma$ of the potential (\ref{u-true-bosons}), obtained 
by solving numerically the integral equation (\ref{bosonic-t-matrix-1}) (asterisks) 
and by the approximate relation (\ref{t-0-approx}) (broken line). 
The full line represents the Born approximation $a/a_{F} = 2 \gamma$.}
\label{fig:avsga}
\end{figure}

The results for $a$ (in units of $a_{F}$) vs. $\gamma$ are shown in 
Fig.~\ref{fig:avsga}, where the asterisks corresponds to the numerical solution 
of Eq.~(\ref{bosonic-t-matrix-1}) and the full curve represents the Born approximation 
$a^{\mathrm{Born}}/a_{F} = 2 \gamma$.
Note that, in the region of interest ($\gamma \approx 1$), 
$a^{\mathrm{Born}}/a \approx 3$ and the solution of Eq.~(\ref{bosonic-t-matrix-1}) 
cannot be obtained by perturbative methods (the perturbative region - where the Born 
series associated with the integral equation (\ref{bosonic-t-matrix-1}) for a repulsive 
potential would converge - is, in fact, limited by $a^{\mathrm{Born}}/a \lapprox 2$ 
and corresponds to $\gamma \lapprox 0.5$).
For this reason, any truncation of the integral equation would not be justified.
Note also that the difference between $a^{\mathrm{Born}}$ and $a$ increases drastically 
as $\gamma$ increases (we have verified that $a/a_{F}$ is proportional to 
$\log{\gamma}$ at least over eight decades).

Reducing the integral equation to a matrix equation (by discretizing the 
integral it contains) may, however, be less accurate when solving the original 
equation (\ref{bosonic-t-matrix-0}). 
In that case, it might be useful to make comparison with an alternative 
approximate result that avoids matrix inversion altogether.
The solution of the two-body problem (\ref{bosonic-t-matrix-1}) suggests us to get
this approximate result along the following lines.
Quite generally, suppose that $u({\mathbf q})$ has a characteristic range $q_{o}$. 
It then follows from Eq.~(\ref{bosonic-t-matrix-1}) that 
$t({\mathbf q},0)$ has the \emph{same range\/} $q_{o}$ and is \emph{proportional to\/} 
$u({\mathbf q})$ for $|{\mathbf q}| \gg q_{o}$ (both properties can be verified 
by working out specific examples). Thus, if we set

\begin{equation}
\frac{t({\mathbf q},0)}{t(0)} \, = \, \lambda({\mathbf q}) \, \frac{u({\mathbf q})}
                                                 {u(0)} \, \, \,  ,     \label{lambda}
\end{equation}

\noindent
then $\lambda({\mathbf q})$ varies slowly from $\lambda(0) = 1$ to $\lambda(\infty) < 1$
(for a repulsive potential).
Since the integral in Eq.~(\ref{bosonic-t-matrix-1}) weighs more ``small'' than
``large'' wave vectors, we can replace $\lambda({\mathbf q}) \simeq \lambda(0) = 1$
therein and obtain

\begin{equation}
t(0) \, \simeq \, \frac{u^{2}(0)}{u(0) + \int\!\frac{d{\mathbf q}}{(2 \pi)^{3}}
 \frac{u^{2}({\mathbf q})}{{\mathbf q}^{2}/M}}    \,\, \, .          \label{t-0-approx}
\end{equation}

\noindent
The values of $a/a_{F}$ obtained with this approximation are reported in
Fig.~\ref{fig:avsga} (broken line), and show remarkable agreement with the full 
numerical solution of the integral equation (\ref{bosonic-t-matrix-1}), from the 
perturbative region ($\gamma \ll 1$) up to the region  of interest ($\gamma \approx 1$).

The same sort of approximation can be adopted when solving the original integral 
equation (\ref{bosonic-t-matrix-0}) for composite bosons. 
In this case, we set approximately

\begin{equation}
\frac{\bar{t}_{B}(q,-q,0,0)}{\bar{t}_{B}(0)} \, \simeq \, 
\frac{\bar{u}_{2}(q,-q,0,0)}{\bar{u}_{2}(0)}                    \label{tB-vs-u2-approx}
\end{equation}

\noindent
and obtain from Eq.~(\ref{bosonic-t-matrix-0}) the expression

\begin{equation}
\bar{t}_{B}(0) \, \simeq \, \frac{\bar{u}_{2}(0)^{2}}{\bar{u}_{2}(0) + 
\frac{1}{\beta {\mathcal V}} \, \sum_{q} \, \bar{u}_{2}(q,-q,0,0)^{2} \,
\Gamma_{o}(q) \, \Gamma_{o}(-q)}     \,\, \, .                   \label{tB-0-approx}       
\end{equation}

The calculation of $\bar{u}_{2}(q,-q,0,0)$ according to the general expression
(\ref{two-body-potential-bar}) can be reduced to the numerical evaluation of 
an integral over $|{\mathbf k}|$, with $\mu = - \epsilon_{o}/2$ in the fermionic
single-particle Green's functions.
The frequency part of the sum over $q$ in Eq.~(\ref{tB-0-approx}) can further be
reduced to an integral, since for vanishing density the range of temperature
we are interested in vanishes.
An accurate numerical evaluation of the ensuing integral over $|{\mathbf q}|$ and
$\Omega$ in Eq.~(\ref{tB-0-approx}) yields

\begin{equation}
\frac{a_{B}}{a_{B}^{\mathrm{Born}}} \, = \, \frac{\bar{t}_{B}(0)}{\bar{u}_{2}(0)} 
\, \simeq  \, \frac{1}{2.69}                                       \label{1-2.69}
\end{equation}

\noindent
within an estimated $1\%$ numerical accuracy.

Full numerical calculation of Eq.~(\ref{bosonic-t-matrix-0}) requires 
us to introduce a finite-size mesh for the variables ($|{\mathbf q}|$, $\Omega$) as
well as ($|{\mathbf q}'|$, $\Omega'$), with the angular integral over $\hat{q'}$
affecting only the function $\bar{u}_{2}(q,-q,q',-q')$.
Equation (\ref{bosonic-t-matrix-0}) is thus reduced to a set of coupled equations
for the unknowns $\bar{t}_{B}(|{\mathbf q}|,\Omega;|{\mathbf q}|,-\Omega;0;0)$,
which we have solved by the standard Newton-Ralphson algorithm with a linear
interpolation for the integral over $|{\mathbf q}'|$ and $\Omega'$.
In this way we obtain

\begin{equation}
\frac{a_{B}}{a_{B}^{\mathrm{Born}}} \, = \, \frac{\bar{t}_{B}(0)}{\bar{u}_{2}(0)} 
\, \simeq  \, \frac{1}{2.65}                                       \label{1-2.65}
\end{equation}

\noindent
within an estimated $5\%$ numerical accuracy. The agreement between the results
(\ref{1-2.69}) and (\ref{1-2.65}) implies that the numerical procedures we have 
adopted are accurate enough for the present calculation.

To verify that this result could not be inferred from a perturbative expansion of
the integral equation (\ref{bosonic-t-matrix-0}), we calculate eventually the second 
term on the right-hand side of Eq.~(\ref{bosonic-t-matrix-0}) by replacing
$\bar{t}_{B}(q',-q',0,0)$ therein with $\bar{u}_{2}(q',-q',0,0)$ and by setting
$q=0$ everywhere for convenience. In this way we obtain [cf. Eq.~(\ref{tB-0-approx}]:

\begin{eqnarray}
\bar{t}_{B}(0) & \simeq & \bar{u}_{2}(0) \left[1 - 
\frac{1}{\beta {\mathcal V}} \, \sum_{q'} 
\frac{\bar{u}_{2}(q',-q',0,0)^{2}}{\bar{u}_{2}(0)}\right.
\nonumber\\
&\times&\left. \Gamma_{o}(q') \, \Gamma_{o}(-q') + \cdots \phantom{\frac{1}{1}}\right]   \nonumber \\
& = & \bar{u}_{2}(0) \left( 1 \, - \, 1.69 \, + \, \cdots \right)  \,\, , 
                                                              \label{bos-t-matr-0-pert}
\end{eqnarray}

\noindent
showing clearly that the geometric series would not converge in this case.

To summarize, we have shown that, in the strong-coupling limit, the value
$a_{B} = 2 a_{F}$ obtained for the composite-boson scattering length within
the self-consistent fermionic T-matrix approximation,\cite{Haussmann} is
modified to $a_{B} \simeq (3/4) a_{F}$ by the correct inclusion of \emph{all\/}
low-density contributions for a system of composite bosons.
\section{Concluding remarks}

In this paper, we have determined the correct diagrammatic approximation for a
``dilute'' system of composite bosons, which form as tightly bound pairs of fermions
in the limit of strong attraction between the constituent fermions.
We have emphasized that it is physically the comparison of the average interparticle
distance to the characteristic length associated with  the \emph{residual\/}
interaction between the composite bosons to determine the ``diluteness'' condition
in the strong-coupling limit of the original fermionic attraction.
For this reason, the simple argument, according to which to a ``dilute'' system of
fermions in the weak-coupling limit there necessarily corresponds a ``dilute'' system
of composite bosons in the strong-coupling limit, might be misleading, since two
different interactions actually control the ``diluteness'' condition in the two limits.
It is thus essential to treat the residual interaction between the composite bosons
with the due care, in order to control the strong-coupling limit of the theory
appropriately.
In this context, it is worth mentioning that the importance of a proper treatment
of the residual boson-boson interaction in the strong-coupling limit has been
emphasized in the pioneering paper by Nozi\`{e}res and Schmitt-Rink,\cite{NSR}
but never duly taken into account in subsequent work.

We have also shown that the selection of the diagrammatic contributions according to 
the ``diluteness'' parameter proceeds along quite different lines in the weak-coupling
limit (where the small parameter is $k_{F} a_{F}$) and in the strong-coupling limit
(where the small parameter is $\rho_{B}^{1/3} a_{B}$).\cite{footnote-1}
Accordingly, diagrammatic contributions of the \emph{same\/} order in 
$\rho_{B}^{1/3} a_{B}$ in the strong-coupling limit correspond, in general, to 
\emph{different\/} powers of $k_{F} a_{F}$ in the weak-coupling limit.
Mathematically, this difference is due to the behaviour of the fermionic
particle-particle ladder, which reduces to a constant in the weak-coupling limit
but develops a singularity in the strong-coupling limit.

Our selection of diagrammatic contributions has rested on a suitable regularization 
of the fermionic interaction, which has caused the ratio between the
particle-particle and particle-hole contributions to be infinite.
For a Hubbard Hamiltonian on a lattice, where this regularization cannot be applied, 
we expect the difference between p-p and p-h contributions to be less extreme albeit 
still appreciable, so that our selection of diagrammatic contributions may still 
remain valid.\cite{footnote-Perali}

Quite generally, we have remarked that, with our choice of the fermionic interaction,
the most general structure of the diagrammatic theory is constructed with the ``bare''
particle-particle ladder plus an infinite set of (four, six, $\cdots$,-point) vertices.
This remains true for \emph{any\/} value of the fermionic coupling and not just in
the strong-coupling limit where the composite bosons form.
We have also remarked that the ``diluteness'' parameter ($k_{F} a_{F}$ or
$\rho_{B}^{1/3} a_{B}$) emerges \emph{naturally\/} from the theory, both in the
weak- and strong-coupling limits, without having to be imposed as an external
condition.
Accordingly, keeping tracks of the powers of this small parameter in the
diagrammatic theory can be relevant \emph{only\/} in the weak- and strong-coupling
limits.
In the intermediate-coupling (crossover) region, on the other hand, a small parameter
is lacking and consequently the diagrammatic approximations cannot be controlled
by any means.\cite{footnote-QMC}

For these reasons, implementing the self-consistency of the fermionic Green's functions
within the fermionic T-matrix approximation does not seem \emph{a priori\/} to be an 
important issue for the BCS-BE crossover.
Self-consistency, in fact, drops out in the weak- and strong-coupling limits when
the ``diluteness'' parameter is small, while in the intermediate (crossover) region
inclusion of self-consistency within the fermionic T-matrix approximation
(as well as within any other approximation over and above it) \emph{cannot anyway be 
controlled\/} by the lack of a small parameter (even though inclusion of 
self-consistency might produce in practice sizable numerical effects).

We have emphasized in this paper that the (self-consistent) fermionic T-matrix 
approximation does not account properly for the boson-boson interaction in the 
strong-coupling limit, at least in \emph{three\/} dimensions.
This approximation, however, has been recently adopted to discuss pseudo-gap and 
related issues within the negative-$U$ Hubbard model in \emph{two\/} 
dimensions.\cite{Serene,EN,Kyung}
Assessing to what extent the approach we have developed in this paper can be carried 
over to the two-dimensional case is not \emph{a priori\/} evident and will require
further investigations.
From physical intuition one would expect the bosonic regime to be reached even more 
effectively in two than in three dimensions, insofar as the two-fermion bound state 
is present in two dimensions for any (attractive) coupling strength.\cite{Randeria-89}
Our dealing with the three-dimensional case first was required for manifesting at the
outset the effects on the BCS-BE crossover due to the progressive formation of 
bound-fermion pairs, thus isolating them from other effects which are peculiar to the 
two-dimensional case.

We have considered in this paper temperatures \emph{above\/} $T_{c}$ only.
Extending our approach below $T_{c}$ even for the three-dimensional case will 
also require further investigations. 
In fact, divergencies occur in the particle-particle
channel when approaching the critical temperature,\cite{footnote-mu}
over and above the divergencies due to the formation of a bound state in the 
associated two-body problem, which we have taken care of in this paper.

It is interesting to point out that, when written in terms of the constituent
fermions (cf.\ Fig.~\ref{fig:AppB2} below), the diagrammatic structure introduced by
our approach bears some analogy with the parquet approximation utilized for the
positive-$U$ Hubbard model.\cite{Janis}
In addition, the level of numerical effort required to solve our integral equation
Eq.~(\ref{bosonic-t-matrix}) for the generalized T-matrix for composite bosons
appears to be comparable with the effort required to solve the integral equations
defining the parquet scheme.\cite{Janis-2}

It is also interesting to point out the strong analogy between the present
treatment of the BCS to BE crossover in a condensed matter system and the 
so-called
OAI mapping introduced some time ago in nuclear physics,\cite{Iach1,Iach2}
where a systematic mapping between the diagrammatic theories for (composite)
bosons and (constituent) fermions is also provided, albeit in a quite different
physical context and with the use of approximations more specific to the 
nuclear problem.
 
A final comment is in order about the emphasis we have given to the correct
description of the boson-boson interaction in the strong-coupling limit.
We know, in fact, that in the \emph{extreme\/} strong-coupling limit the
boson-boson interaction can be neglected and the composite bosons become
effectively free.
Since this extreme limit of free composite bosons is also retrieved
by the (self-consistent) fermionic T-matrix approximation, it would appear that the 
improvement brought about by our theory (which takes specifically into account the 
interaction between the composite bosons) affects only \emph{the way this extreme limit 
is approached\/}.
That the important physics may reside in the correct description of the 
\emph{approach\/} to the asymptotic regimes and not merely of the asymptotic 
regimes themselves, can be confirmed from previous experience with the 
so-called Hubbard-III solution;\cite{Hubbard-III} in that case, neither the 
weakly-interacting (Fermi liquid) nor the strongly-interacting (Kondo) physics
is reproduced, although the extreme (non-interacting and atomic) limits are
both retrieved.
\acknowledgments

We are indebted to C. Castellani, C. Di Castro, M. Grilli, F. Iachello, 
V. Jani\v{s}, A. Perali, and F. Pistolesi for helpful discussions.
One of us (P.P.) gratefully acknowledges receipt of a postdoctoral research 
fellowship from the Italian INFM under contract PRA-HTCS/96-98.

\appendix
\section{Composite-boson propagator and two-fermion Green's function}

It was shown in Ref.\ 16 (cf.\ Appendix A therein) that the average value of 
$b(q=0)$ (taken with the effective bosonic action (\ref{eff-action})) is proportional
to the average value of the operator (\ref{quadratic-operator}) (taken with the
original fermionic action (\ref{action})). 
Purpose of this Appendix is to show that a similar relation holds between the ``full'' 
bosonic propagator $<b^*(q) b(q)>_{S_{\mathrm{eff}}}$ and the average value of the 
product of two operators (\ref{quadratic-operator}).

To this end, it is convenient to add to the original fermionic action (\ref{action})
the following (source) term

\begin{equation}
\delta S \, = \, \int_0^\beta d\tau \sum_{{\mathbf q}}  \left[ 
J({\mathbf q},\tau) \bar{{\mathcal B}}({\mathbf q},\tau) +
J^*({\mathbf q},\tau)      {\mathcal B}({\mathbf q},\tau)  \right]    \label{A-source}
\end{equation}

\noindent
and to define the fermionic generating functional [cf. Eq.~(\ref{partition-function})]

\begin{equation}
{\mathcal Z}[J,J^*] \, = \, 
\frac{\int {\mathcal D}\bar{c} {\mathcal D}c \,\, \exp\{-S - \delta S\}}
{\int {\mathcal D}\bar{c} {\mathcal D}c \,\, \exp\{-S\}} \, .        \label{A-gen-funct}
\end{equation}

\noindent
Introducing the Hubbard-Stratonovich transformation (\ref{Hub-Straton}), shifting
the bosonic integration variables by letting
$b({\mathbf q},\tau) \rightarrow b({\mathbf q},\tau) \, + \, J({\mathbf q},\tau)$,
and integrating out the Grassmann variables, we obtain the exact identity:

\begin{eqnarray}
\left. \frac{\delta {\mathcal Z}[J,J^*]}
{\delta J({\mathbf q},\tau ') \delta J^*({\mathbf q},\tau)} 
\right|_{J=J^*=0}&=& 
\frac{1}{V^2} <b^*({\mathbf q},\tau ')  
b({\mathbf q},\tau)>_{S_{\mathrm{eff}}}\nonumber\\
&+&  
\frac{1}{V} \delta (\tau - \tau ') \, .                         \label{A-diff}
\end{eqnarray}

\noindent
Upon differentiating the original expression (\ref{A-gen-funct}) in a similar way 
and comparing with Eq.~(\ref{A-diff}), we obtain eventually the desired relation:

\begin{eqnarray}
<b^*({\mathbf q},\tau ')  b({\mathbf q},\tau)>_{S_{\mathrm{eff}}} &=& V^{2} 
<\bar{{\mathcal B}}({\mathbf q},\tau ')  {\mathcal B}({\mathbf q},\tau)>_{S}
\nonumber\\
&-& 
V \, \delta (\tau - \tau ') \, ,                                \label{A-gen-tau}
\end{eqnarray}

\noindent
where the last term on the right-hand side can be dropped since it vanishes with 
the regularization (\ref{vo}) in the limit $k_{o} \rightarrow \infty$.

\begin{figure}
\narrowtext
\epsfxsize=3.3in 
\epsfbox{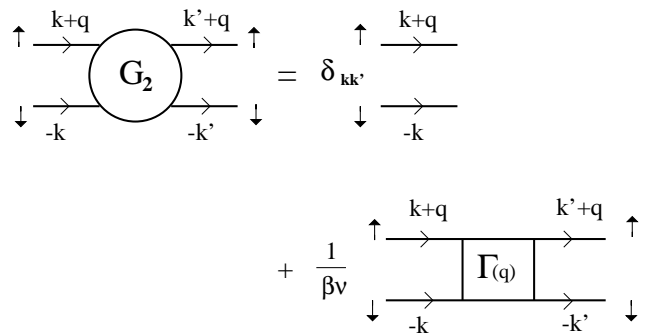}
\vspace{.2truecm}
\caption{Fermionic two-particle Green's function ${\mathcal G}_{2}$,
expressed in terms of the ``full'' single-particle Green's function ${\mathcal G}$ 
(represented here by a line) and of the generalized particle-particle 
ladder $\Gamma$ 
(represented by a box). The factor $(\beta {\mathcal V})^{-1}$ originates from the 
diagrammatic rules. Spin labels are indicated by up and down arrows.}        
\label{fig:AppA}
\end{figure}

The result (\ref{A-gen-tau}) can be conveniently interpreted in diagrammatic terms
as follows.
Taking the Matsubara Fourier transform of both sides of Eq.~(\ref{A-gen-tau})
yields

\begin{eqnarray}
& &<b^*(q)  b(q)>_{S_{\mathrm{eff}}}=V^{2} \sum_{k,k'}w\left({\mathbf k} + {\mathbf q}/2\right) w\left({\mathbf k'} + {\mathbf q}/2\right)
\nonumber\\
& &\times<\bar{c}_{\uparrow}(k+q) \bar{c}_{\downarrow}(-k)
     c_{\downarrow}(-k') c_{\uparrow}(k'+q)>_{S}\nonumber\\
& &\equiv V^{2} \, \sum_{k,k'} \, 
w\left({\mathbf k} + {\mathbf q}/2\right) w\left({\mathbf k'} + 
{\mathbf q}/2\right) {\mathcal G}_{2}(k,k';q)                                   \label{A-gen-omega}
\end{eqnarray}

\noindent
which defines the fermionic two-particle Green's function ${\mathcal G}_{2}$ in the 
particle-particle channel. 
This function is depicted in Fig.~\ref{fig:AppA}, in terms of the single-particle 
fermionic Green's functions ${\mathcal G}$ and of the generalized particle-particle 
ladder $\Gamma$.
Note that the external single-particle lines of Fig.~\ref{fig:AppA} (as well as 
the single-particle lines contained in $\Gamma$) include, in principle, self-energy 
insertions.

Owing to our choice of a ``contact'' potential, the generalized particle-particle
ladder of Fig.~\ref{fig:AppA} depends only on the sum $q$ of the incoming (outgoing) 
four-momenta.
For this reason, the sums over $k$ and $k'$ in Eq.~(\ref{A-gen-omega}) are decoupled
and affect only the external single-particle lines. Let

\begin{equation}
{\mathcal I}_{pp}(q) \, = \, \frac{V}{\beta} \, \sum_{k}
w\left({\mathbf k} + {\mathbf q}/2\right) \, {\mathcal G}(k+q) {\mathcal G}(-k)
                                                               \label{A-GG}
\end{equation}

\noindent
be the analogue of the quantity $I_{pp}(q)$ introduced in  Eq.~(\ref{I-pp}),
where now the full single-particle fermionic Green's function ${\mathcal G}$ 
replaces its bare counterpart ${\mathcal G}^{o}$.
We know already that $I_{pp} = - 1 + {\mathcal O}(k_{o}^{-1})$, whereby only 
the
behavior of ${\mathcal G}^{o}$ for large wave vectors contributes to the finite term.
The self-energy insertions contained in ${\mathcal G}$ cannot modify this result, 
since they vanish for large wave vectors (provided a well-defined \emph{finite\/}
expression is associated with the self-energy itself in the limit 
$k_{o} \rightarrow \infty$). 
This remark implies that also ${\mathcal I}_{pp} = - 1 + 
{\mathcal O}(k_{o}^{-1})$.

Combining Eq.~(\ref{A-gen-omega}) with the representation of Fig.~\ref{fig:AppA}, 
we then obtain the following relation between the ``full'' bosonic propagator and 
the generalized particle-particle ladder:

\begin{equation}
<b^*(q)  b(q)>_{S_{\mathrm{eff}}} \, = \, \frac{\beta}{{\mathcal V}} \,\, 
\Gamma(q)
\label{A-final}
\end{equation}

\noindent
of which Eq.~(\ref{bare bb}) of the text is the simplest approximation.
It is clear from our derivation that the result (\ref{A-final}) holds 
irrespective of the value of the fermionic scattering length $a_{F}$.

The equivalence (\ref{A-final}) between the generalized particle-particle ladder
and the ``full'' bosonic propagator is basic to the purposes of the present paper,
since it enables us to interpret in a rigorous fashion the general diagrammatic 
structure of $\Gamma(q)$ in terms of the building blocks of the diagrammatic structure 
of $<b^*(q)  b(q)>_{S_{\mathrm{eff}}}$, namely, the ``bare'' bosonic propagator
$<b^*(q)  b(q)>_{S^{2}_{\mathrm{eff}}}$ given by Eq.~(\ref{bare bb}) and the four, six, 
$\cdots$, -point interaction vertices given by Eqs.~(\ref{two-body-potential}), 
(\ref{three-body-potential}), $\cdots$, in the order.
\section{Fermionic diagrams corresponding to the T-matrix approximation
for composite bosons}

In this Appendix we discuss the mapping between the diagrammatic structures of the
composite bosons and the constituent fermions, for the case when only the four-point
interaction vertex for composite bosons is retained and all higher-order vertices are
neglected.
In particular, we shall consider the class of diagrams of interest depicted in
Fig.~\ref{fig:comboslo}b, corresponding to the T-matrix approximation
for composite bosons. 
To make our argument more complete, we shall also consider one additional bosonic 
diagram not belonging to this class, owing to its peculiar topological structure.
By working out these two examples in detail, we shall establish a definite 
correspondence between the value of the symmetry factors for the bosonic diagrammatic 
structure and the number of independent diagrams in the associated fermionic structure,
as anticipated in the Introduction.
Besides providing a compelling check on the general structure of our mapping,
determining the \emph{explicit form\/} of the fermionic diagrams proves useful to 
verify that our generalized T-matrix approximation correctly reduces to the 
standard fermionic (Galitskii) approximation in the weak-coupling limit.

We begin by determining the symmetry factors associated with the diagrams of
Fig.~\ref{fig:bosloop}b for true bosons, where the two-body interaction is
assumed to be suitably symmetrized.
The symmetry factor ${\mathcal S}_{L}$ associated with the diagram containing $L$
interaction vertices is given by the ratio of the number of ways to construct this
diagram from the corresponding prediagram and the factor $4^{L} L!$ originating
from the perturbative expansion at order $L$.\cite{Popov-1,Popov-2}
By definition, in the prediagram only the external (incoming and outgoing) arrows 
of the self-energy and the interaction vertices appear, but not the propagators 
that join the vertices among themselves or with the external arrows.
Note that, for a complex bosonic field, the interaction vertices bear directional
arrows and, consequently, the number of ways to join them with the propagators 
differs from the value of the standard $\phi^{4}$ theory with a real field.
One can readily show that the number of ways to construct the diagram of 
Fig.~\ref{fig:bosloop}b with $L$ interaction vertices from the corresponding
prediagram is given by $2^{L+1} L!$, yielding for the symmetry factor the value

\begin{equation}
{\mathcal S}_{L} \, = \, \frac{2^{L+1} L!}{4^{L} L!} \, = \, \frac{1}{2^{L-1}} \, \, .
                                                         \label{B-symmetry-factor}
\end{equation}

\noindent
Since for this set of diagrams all arrangements of vertices and propagators produce 
different contractions (according to Wick's theorem), no other factor is required in 
this case beside the symmetry factor (\ref{B-symmetry-factor}).\cite{BDFN}

It is clear that with the diagrams of Fig.~\ref{fig:comboslo}b for composite bosons is 
also associated the \emph{same\/} symmetry factor (\ref{B-symmetry-factor}), as they 
share the same topological structure of the diagrams of Fig.~\ref{fig:bosloop}b.
Since the symmetry factor of any fermionic diagram is instead bound to be unity,\cite{FW} 
we expect ${\mathcal S}_{L}^{-1}$ \emph{identical\/} fermionic diagrams to be 
associated with a given diagram for composite bosons having symmetry factor 
${\mathcal S}_{L}$.

\begin{figure}
\narrowtext
\epsfysize=4.0in\hspace{0.5in} 
\epsfbox{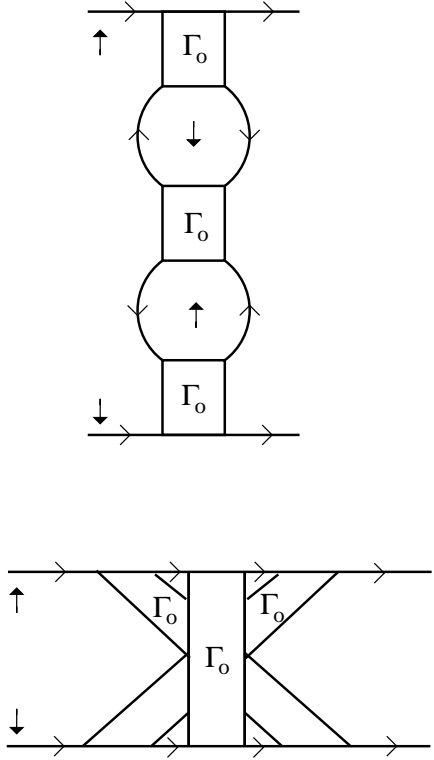}
\vspace{.2truecm}
\caption{The two distinct fermionic diagrams corresponding
to the bosonic diagram with two interaction vertices ($L=2$ and ${\mathcal S}_{2}=1/2$) 
depicted in Fig.~\ref{fig:comboslo}b.
Boxes and full lines stand for the ``bare'' particle-particle ladder $\Gamma_{o}$ 
and for the ``bare'' single-particle fermionic Green's function ${\mathcal G}^{o}$, 
respectively.}         
\label{fig:AppB1}
\end{figure}

To verify this statement, we recall the correspondence (shown in
Fig.~\ref{fig:cobos}) of the propagator and vertex of the bosonic theory 
with the building blocks of the fermionic diagrammatic structure (which takes also
into account the spin labels).
In subsection IIB we have already discussed the case with $L=1$ and 
${\mathcal S}_{1}=1$, 
with which two distinct fermionic diagrams have consistently been associated 
(cf. Fig.~\ref{fig:bosHF}).
The case with $L=2$ and ${\mathcal S}_{2}=1/2$ is shown graphically in
Fig.~\ref{fig:AppB1}. 
In this case, it can be readily verified that the four possible fermionic 
diagrams (obtained by associating \emph{two\/} distinct fermionic structures 
to each bosonic 
vertex via the rules of Fig.~\ref{fig:cobos}b) are equal in pairs.
The resulting factor of 2 multiplying each independent fermionic diagram is 
then canceled by the symmetry factor ${\mathcal S}_{2}=1/2$, so that each 
independent fermionic diagram is correctly multiplied by unity, as anticipated.

By a similar token, when $L=3$ one can verify that, out of the eight possible fermionic
diagrams, only two are independent. The ensuing factor of 4 in front of each of these
two independent diagrams is then canceled by the symmetry factor 
${\mathcal S}_{3}=1/4$.
More generally, if we call ``up'' and ``down'', respectively,  the two fermionic 
structures associated in Fig.~\ref{fig:cobos}b with the bosonic four-point vertex,
it can be verified that fermionic diagrams, obtained from each other by replacing a
\emph{pair\/} of ``up'' by a pair of ``down'' fermionic structures, are equal.
This property is \emph{sufficient\/} to guarantee that each distinct fermionic diagram 
is correctly multiplied by unity.
Consider, for instance, two nontrivial cases with odd ($L=5$) and even 
($L=6$) number of bosonic vertices.
When $L=5$, $2^{5}=32$ fermionic diagrams are generated from the bosonic diagram.
In this case, from the single fermionic diagram with all ``up'' structures we can  
generate 5!/(3!2!)=10 diagrams with three ``up'' and two ``down'' structures, and
5!/(1!4!)=5 diagrams with one ``up'' and four ``down'' structures, for a total of
1+10+5=16 identical fermionic diagrams. 
The same number of identical diagrams results from the other independent diagram 
having all ``down'' structures (which, evidently, cannot be obtained from the 
previous diagram with all ``up'' structures by replacing an even number of ``up'' 
structures by ``down'' structures).
The resulting factor of 16 associated with each of these two independent fermionic 
diagrams is then canceled by the symmetry factor ${\mathcal S}_{5}=1/16$, and each 
independent fermionic diagram is correctly multiplied by unity.

When $L=6$, $2^{6}=64$ fermionic diagrams are generated from the bosonic diagram.
In this case, the two independent diagrams (out of the 64 possible fermionic diagrams)
have either all ``up'' structures or five ``up'' and one ``down'' structures.
To the first case we associate 1+6!/(4!2!)+6!/(2!4!)+1=32 identical diagrams, and
to the second case 6!/(5!1!)+6!/(3!3!)+6!/(1!5!)=32 identical diagrams.
The resulting factor of 32 multiplying each independent fermionic diagram is 
then canceled by the symmetry factor ${\mathcal S}_{6}=1/32$, and each independent 
fermionic diagram is again correctly multiplied by unity.

\begin{figure}
\narrowtext
\epsfysize=4.0in 
\epsfbox{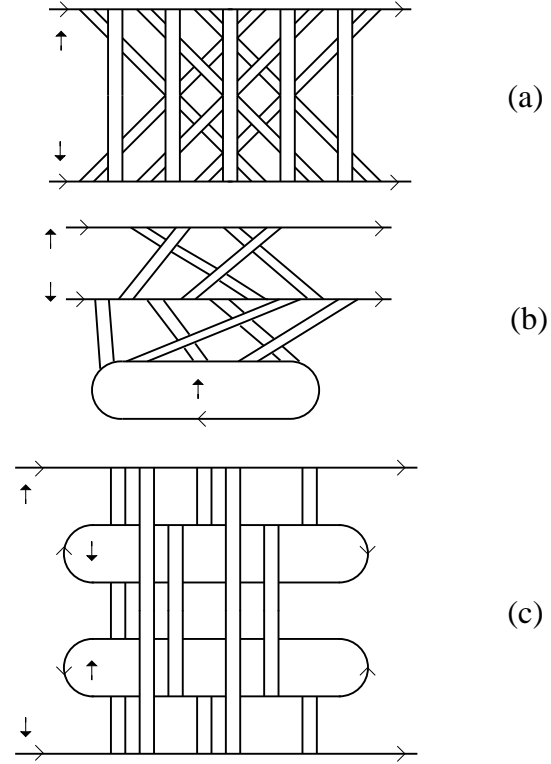}
\hspace{0.2in}
\vspace{.2truecm}
\caption{Representative diagrams of the three
classes of fermionic diagrams associated with the bosonic T-matrix, containing 
\emph{(a)} zero, \emph{(b)} one, and \emph{(c)} two fermionic loops.
Conventions are as in Fig.~\ref{fig:AppB1}.}         
\label{fig:AppB2}
\end{figure}

Quite generally, one can show that the fermionic diagrams associated with the
T-matrix diagrams of Fig.~\ref{fig:comboslo}b for composite bosons can 
be organized into \emph{three classes\/}, containing zero, one, and two fermionic 
loops, respectively.
A representative diagram of each of these three classes is depicted in 
Fig.~\ref{fig:AppB2}a, Fig.~\ref{fig:AppB2}b, and Fig.~\ref{fig:AppB2}c, 
in the order.
Note that fermionic diagrams containing zero or two loops are associated with
bosonic diagrams with an \emph{even\/} number of interaction vertices, while
fermionic diagrams containing one loop are associated with bosonic diagrams with
an \emph{odd\/} number of interaction vertices (cf. the comment made in 
subsection IIB while discussing Fig.~\ref{fig:bosHF}).

The above analysis can be made more systematic, by allowing also for the possibility
of ``twisting'' pairs of fermionic lines while connecting the fermionic 
particle-particle ladder of Fig.~\ref{fig:cobos}a with the four-point vertex of 
Fig.~\ref{fig:cobos}b.
In other words, one may consider the situation where two pairs of fermionic 
lines, which would not be allowed to connect with each other ``directly'' as their 
spins would not match, are instead allowed to connect by ``twisting'' one pair of 
lines with respect to the other one. 
A careful analysis of the unraveling of the fermionic diagrams containing ``twistings'',
however, shows that no new diagram is introduced in this way, over and above the 
diagrams depicted in Fig.~\ref{fig:AppB2} which have been generated by not 
allowing ``twisted'' connections.

We are now in a position to verify that the fermionic diagrams, associated with
the generalized T-matrix approximation for composite bosons, correctly reduce to the 
standard Galitskii approximation in the weak-coupling limit, in the sense that they 
yield contributions of progressively higher order in the small parameter $k_{F} a_{F}$
with respect to the ``bare'' particle-particle ladder.
For instance, while the particle-particle ladder is proportional to $a_{F}$ in the
weak-coupling limit (cf. Eq.~(\ref{pp-wc})), corrections to this ladder due to the
simple diagrams of Fig.~\ref{fig:bosHF} are readily seen to be smaller by a factor 
$(k_{F} a_{F})^{2}$.
Similarly, corrections to the particle-particle ladder due to the diagrams of
Fig.~\ref{fig:AppB1} are smaller by a factor $(k_{F} a_{F})^{4}$ with respect to
the ``bare'' ladder.
In general, to estimate in the weak-coupling limit the order in $k_{F} a_{F}$
of a given diagram contributing to the generalized particle-particle ladder 
$\Gamma$, we can rely on \emph{dimensional considerations\/}, by counting the powers 
of $a_{F}$ in terms of the number of ``bare'' ladders $\Gamma_{o}$ and the powers of
$k_{F}$ in terms of the dimensionality of the four-vector sums over products of 
internal single-particle fermionic Green's functions (which can safely be done because 
all sums are convergent).
By this procedure, we obtain that, in the weak-coupling limit, the diagram of order 
$L$ within the bosonic T-matrix approximation is smaller by a factor 
$(k_{F} a_{F})^{2L}$ with respect to the ``bare'' ladder $\Gamma_{o}$.

This proves that diagrams of the same order in $\rho_{B}^{1/3} a_{B}$ in the bosonic
(strong-coupling) limit end up having \emph{different} orders in $k_{F} a_{F}$ 
in the fermionic (weak-coupling) limit; accordingly, they would have been dismissed
as being irrelevant, if the selection of diagrams would have been made directly for 
the weak-coupling limit.

\begin{figure}
\narrowtext
\epsfxsize=2.8in 
\epsfbox{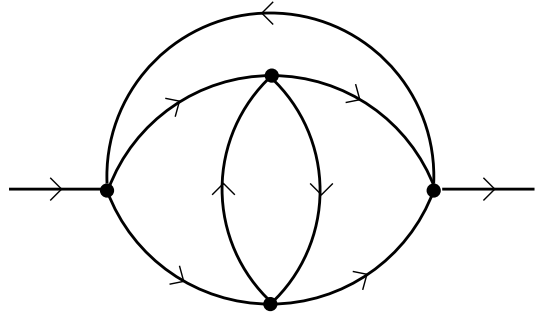}
\vspace{.2truecm}
\caption{Bosonic diagram with four interaction vertices, 
which does not belong to the set of T-matrix diagrams and possesses a special
internal symmetry.}         
\label{fig:AppB3}
\end{figure}

Finally, let us consider the bosonic diagram of Fig.~\ref{fig:AppB3} which does
not belong to the set of T-matrix diagrams.
For this diagram, a naive counting of the number of ways it can be constructed from 
the corresponding prediagram would give unity for the symmetry factor.
However, a more careful analysis shows that a permutation of the two internal vertices 
is actually equivalent to a renaming the propagators joining these vertices,
and has thus not to be regarded as an independent permutation.\cite{BDFN}
This implies that the bosonic symmetry factor for the diagram of Fig.~\ref{fig:AppB3}
is 1/2 and not 1.
As a matter of fact, when generating the fermionic diagrams associated with this
bosonic diagram according to the correspondence rules of Fig.~\ref{fig:cobos},
one finds that the $2^{4}=16$ diagrams (produced by the presence of four bosonic 
interaction vertices) are equal in pairs, yielding just 8 distinct fermionic diagrams 
in agreement with the value 1/2 of the bosonic symmetry factor.
As before, this result is not affected by introducing ``twistings''.

Although we have examined in detail only two kinds of bosonic diagrams, we
expect the correspondence between the bosonic symmetry factor and the number of
independent fermionic diagrams to remain valid for all possible diagrams. 



\begin{references}
\bibitem{footnote-1} In the \emph{extreme\/} strong-coupling limit, the interaction 
between the composite bosons is bound to vanish (cf. \cite{Haussmann,PS-96}). 
Physically, this is due to the fact that, for a point-contact interaction between 
the constituent fermions, the composite bosons of size $a_{F}$ mutually interact 
only when overlapping.
For this reason, the bosonic scattering length $a_{B}$ should be proportional to
$a_{F}$. The condition $k_{F} a_{F} \ll 1$ can then be interpreted as 
$\rho_{B}^{1/3} a_{B} \ll 1$, which is the ``diluteness'' condition for a Bose gas.
The ``diluteness'' condition thus emerges naturally both in the weak- and 
strong-coupling limits, and should \emph{not\/} be imposed as an additional 
condition on the system.
\bibitem{footnote-3} In principle, by this procedure one should complement the
chosen set of two-fermion diagrams with all companion diagrams, as prescribed by
the Baym-Kadanoff method. In practice, however, these additional diagrams will
not be included in our theory since they are of higher order in the ``gas parameter''
$\rho_{B}^{1/3} a_{B}$. This point will be discussed further in Section III.
\bibitem{footnote-5} It can be readily verified that, if at least one ($2l$)-point
vertex enters a bosonic self-energy diagram, then there are at least ($l-1$) bosonic
cycles in that diagram (see also Ref.\ 33).
Since each bosonic cycle gives a contribution proportional to the density $\rho$
(as shown in Section III), four-point vertices only contribute to the bosonic 
self-energy to lowest order in $\rho$.
\bibitem{footnote-6} Self-consistency gives contributions at least of order 
$(k_{F} a_{F})^{2}$ to the fermionic self-energy. For this reason, it was not
considered by Galitskii in the calculation of the self-energy to order $k_{F} a_{F}$.
A first round of self-consistency was, however, included by Galitskii in the
calculation of the self-energy of a ``dilute'' Fermi gas to order 
$(k_{F} a_{F})^{2}$.\cite{Galitskii}
In the context of the BCS-BE crossover, the complete self-consistency within the 
fermionic T-matrix approximation has been considered only recently 
(cf.\ Refs.\ 8 and 9), and since then utilized by several 
authors in this field.\cite{Micnas,Randeria-97-1,KKK}
\end{references}
\end{document}